\documentclass[prx,reprint,twocolumn,footinbib,longbibliography,superscriptaddress]{revtex4-1}
\usepackage{amsmath}
\usepackage{amsfonts}
\usepackage{amssymb}
\usepackage{times}
\usepackage[dvipdfmx]{graphicx}
\usepackage[usenames,dvipsnames]{xcolor}
\usepackage{bm}
\usepackage{amsthm}
\usepackage{physics}
\usepackage{mathrsfs}

\usepackage{hyperref}

\newcommand{\bmnu}{{\bm \nu}}

\begin{document}
\
\title{Geometrical aspects of entropy production in stochastic thermodynamics based on Wasserstein distance}
\date{\today}

\author{Muka Nakazato}
\affiliation{Department of Physics, The University of Tokyo, 7-3-1 Hongo, Bunkyo-ku, Tokyo 113-0031, Japan}
\author{Sosuke Ito}
\affiliation{Department of Physics, The University of Tokyo, 7-3-1 Hongo, Bunkyo-ku, Tokyo 113-0031, Japan}
\affiliation{Universal Biology Institute, The University of Tokyo,7-3-1 Hongo,  Bunkyo-ku, Tokyo 113-0033, Japan}
\affiliation{JST, PRESTO, 4-1-8 Honcho, Kawaguchi, Saitama, 332-0012, Japan}

\begin{abstract}
We study a relationship between optimal transport theory and stochastic thermodynamics for the Fokker-Planck equation. We show that the lower bound on the entropy production is the action measured by the path length of the $L^2$-Wasserstein distance. Because the $L^2$-Wasserstein distance is a geometric measure of optimal transport theory, our result implies a geometric interpretation of the entropy production. Based on this interpretation, we obtain a thermodynamic trade-off relation between transition time and the entropy production. This thermodynamic trade-off relation is regarded as a thermodynamic speed limit which gives a tighter bound of the entropy production.  We also discuss stochastic thermodynamics for the subsystem and derive a lower bound on the partial entropy production as a generalization of the second law of information thermodynamics.  Our formalism also provides a geometric picture of the optimal protocol to minimize the entropy production. We illustrate these results by the optimal stochastic heat engine and show a geometrical bound of the efficiency.
\end{abstract}

\maketitle

\section{Introduction}
Geometry is a helpful tool to consider the difference between two quantities, and the geometric concept for probability distributions is widely used in statistical physics. For example, the Kullback-Leibler divergence~\cite{cover2012elements} is an information-geometric measure that quantifies the difference between two probability distributions and provides the entropy production~\cite{Jaynes1957information, kawai2007dissipation}. The differential geometry based on the entropy production is regarded as information geometry~\cite{amari2000methods} because the second-order Taylor expansion of the Kullback-Leibler divergence leads to the metric in information geometry called the Fisher metric~\cite{rao1945information}. Based on information geometry, geometric aspects of thermodynamics have been discussed~\cite{weinhold1975metric,ruppeiner1995riemannian,salamon1983thermodynamic,crooks2007measuring, ito2018stochastic1,ito2018unified,nakamura2019reconsideration,parr2020philosophical, aguilera2021unifying}. For example, thermodynamic trade-off relations such as thermodynamic uncertainty relations~\cite{horowitz2019thermodynamic,barato2015thermodynamic,pietzonka2016universal, horowitz2017proof, dechant2018current} and speed limits~\cite{ito2018stochastic1, Shiraishi2018speed, Falasco2020dissipation,yoshimura2020information, gupta2020tighter} have been derived based on the Fisher metric~\cite{ito2018stochastic1,ito2020stochastic,dechant2018multidimensional,hasegawa2019uncertainty,Liu2020thermodynamic, nicholson2020time}. These trade-off relations are mathematically connected~\cite{Pietzonka2018universal,dechant2018multidimensional,Vu2020unified,Yoshimura2021speed}. The Fisher metric also explains the optimal control of thermodynamic systems and their stability~\cite{crooks2012sivak,rotskoff2015optimal,ito2019glansdorff,ashida2020experimental}.

In optimal transport theory~\cite{Villani2003topics, Villani2008Optiaml}, another geometry explains the optimal control and is related to thermodynamics. In optimal transport theory, a geometric measure called the $L^2$-Wasserstein distance quantifies a difference between two probability distributions. A relationship between $L^2$-Wasserstein distance and thermodynamic relaxation has been discussed, especially for the Fokker-Planck equation. For example, R. Jordan, D. Kinderlehrer, and F. Otto showed that the time evolution of the Fokker-Planck equation minimizes the sum of the free energy and the $L^2$-Wasserstein distance~\cite{JKO1998Variational}. A trend to thermodynamic equilibrium for the Fokker-Planck equation has also been discussed using the $L^2$-Wasserstein distance~\cite{Marcowich2000Trend}. Remarkably, the terminology of the entropy production is also used in optimal transport theory~\cite{Villani2003topics}, and a connection between the entropy production and the $L^2$-Wasserstein distance has been discussed~\cite{Villani2008entropy, Villani2000generalization, Arnold2001convex}. Moreover, a relationship between optimal transport theory and information geometry has been mathematically discussed~\cite{amari2018information, amari2019information}.

In parallel with information geometry, optimal transport theory can explain the optimal control of thermodynamic systems and thermodynamic trade-off relations. Optimal transport theory has been used in stochastic thermodynamics to find a heat minimization protocol~\cite{Aurell2011oprimal,Aurell2012Refined, Proesmans2020finite, Proesmans2020optimal}. In the context of a heat minimization protocol, E. Aurell {\it et al.} have derived the lower bound on the entropy production~\cite{Aurell2012Refined} using the Banamou-Brenier formula~\cite{Benamou2000Monge} in optimal transport theory. Recently, A. Dechant and Y. Sakurai have pointed out that this lower bound on the entropy production is regarded as a thermodynamic speed limit~\cite{Dechant2019Wasserstein}. Several thermodynamic trade-off relations about the efficiency of the stochastic heat engine has been derived based on the optimal transport theory~\cite{Fu2021maximal, Van2021geometrical}. Nevertheless the usefulness of the above result, optimal transport theory has not been well focused in the field of stochastic thermodynamics~\cite{sekimoto2010stochastic,seifert2012stochastic}.

This paper shows a novel relationship between optimal transport theory and stochastic thermodynamics for the Fokker-Planck equation. Based on a connection between the entropy production rate and the $L^2$-Wasserstein distance~\cite{Villani2000generalization}, we clarify geometrical aspects of the entropy production and derive several thermodynamic trade-off relations.
To consider an infinitesimal time evolution step, we newly show that the entropy production is bounded by the time integral of the square of the velocity, namely the action in differential geometry, measured by the space of the $L^2$-Wasserstein distance. Furthermore, the entropy production can be proportional to the action with some assumptions where the force is given by the potential. This result provides a geometric interpretation of the entropy production for the Fokker-Planck equation.
Using this geometrical expression of the entropy production, we obtained a lower bound on the entropy production as a generalization of the thermodynamic speed limit, which is tighter than the previous result~\cite{Aurell2012Refined, Dechant2019Wasserstein}. 
Remarkably, the derivation of the new thermodynamic speed limit is the same as the original derivation of the thermodynamic speed limit based on information geometry~\cite{ito2018stochastic1}. Moreover, we discuss stochastic thermodynamics for the subsystem~\cite{parrondo2015thermodynamics,ito2013information,horowitz2014thermodynamics,hartich2014stochastic,horowitz2014secondlaw,ito2015maxwell} and the stochastic heat engine~\cite{Schmiedl2007stochastic} by using the $L^2$-Wasserstein distance. We obtain a tighter bound on the partial entropy production as a generalization of the second law of information thermodynamics~\cite{parrondo2015thermodynamics,ito2013information,horowitz2014thermodynamics,hartich2014stochastic,horowitz2014secondlaw,ito2015maxwell,ito2018unified}. We illustrate our results by using the examples of the harmonic potential where the entropy production is proportional to the action. Based on the geometrical interpretation of the entropy production, we obtain a geometrical constraint of the heat engine's efficiency and analytical derivation of the optimization protocol~\cite{Schmiedl2007optimal} to minimize the entropy production. We also numerically illustrate a tightness of a generalized thermodynamic speed limit and the optimal heat engine based on the Wasserstein distance. 

This paper is organized as follows. In Sec.~II, we review previous results on stochastic thermodynamics and optimal transport theory. We formulate the setup of the Fokker-Planck equation in Sec.~II A. We introduce the concept of the Wasserstein distance in Sec.~II B. 
In Sec.~III, we discuss our main results, which are a geometrical interpretation of the entropy production and new geometric lower bounds on the entropy production. We present a geometrical interpretation of the entropy production in Sec.~III A and discuss new geometric lower bounds on the entropy production in Sec.~III B. In Sec.~IV, we discuss a generalization of the result in Sec.~III for a subsystem. We introduce the setup of a subsystem and generalize the main result in Sec.~IV A. We discuss an information-thermodynamic interpretation and derive a new lower bound on the partial entropy production in Sec.~IV B. In Sec.~V, we illustrate the main result by several examples. In Sec.~V A, we discuss the stochastic heat engine and show a geometric lower bound on the efficiency. In Sec.~V B, we show the optimal protocol to minimize the entropy production based on our geometric interpretation analytically. In Sec.~V C, we numerically illustrate geometric lower bounds on the entropy production and the estimation of the entropy production based on the lower bound. In Sec.~V D, we numerically discuss the optimal protocol to minimize the entropy production for the stochastic heat engine. In Sec.~VI, we conclude this paper with some remarks.

\section{Review on stochastic thermodynamics and optimal transport theory}
\subsection{Stochastic thermodynamics for Fokker-Planck equation}
In this paper, 
we consider the probability distribution $p_t({\bf x})$ 
of a particle in a Euclid $d$-dimensional position ${\bf x} \in X (= \mathbb{R}^d)$ at time $t$. The time evolution of $p_t({\bf x})$ is described by the following Fokker-Planck equation 
for a particle driven by a potential $V_t(x)$ with mobility $\mu$ attached to a heat bath at temperature $T$,
\begin{align}
    \frac{\partial p_t({\bf x})}{\partial t} &= -\nabla \cdot (\bmnu_t ({\bf x})p_t ({\bf x})), \label{fokker-planck}\\
   \bmnu_t ({\bf x}) :&= - \mu \nabla [ V_t ({\bf x})  + T \ln p_t({\bf x}) ],
   \label{fokker-planckassume}
\end{align}
where $\nabla$ is the del oparator, and $\bmnu_t ({\bf x})$ is a quantity called the mean local velocity. We here set the Boltzmann constant to unity $k_{\rm B}=1$. As a continuity equation, the mean local velocity $\bmnu_t ({\bf x})$ is regarded as the velocity field.
In stochastic thermodynamics~\cite{sekimoto2010stochastic}, 
the internal energy $U$, 
the extracted work $dW$, 
the heat received from the heat bath $dQ$, 
and the entropy of the system $S_{\mathrm{sys}}$ at time $t$ are defined as follows,
\begin{align}
    U :&= \int d{\bf x} \ V_t({\bf x}) p_t({\bf x}),  \label{energy}\\
    S_{\mathrm{sys}} :&= - \int d{\bf x} \ p_t ({\bf x}) \ln p_t({\bf x}),  \label{entropy}\\
    \frac{dW}{dt} :&= \int d{\bf x} \  \frac{\partial V_t({\bf x})}{\partial t} p_t({\bf x}),  \label{work}\\
    \frac{dQ}{dt} :&= \int d{\bf x} \ V_t ({\bf x}) \frac{\partial p_t({\bf x})}{\partial t}.
    \label{heat}
\end{align}
By definition, the heat $dQ$ satisfies the first law of thermodynamics $dU/dt = dW/dt + dQ/dt$. From these definitions (\ref{energy})-(\ref{heat}), the entropy production rate at time $t$
\begin{align}
\sigma_t  :=\frac{dS_{\mathrm{sys}}}{dt} - \frac{1}{T} \frac{dQ}{dt}
\end{align}
is calculated as 
\begin{align}
    \sigma_t &= \frac{1}{\mu T}\int d{\bf x} \left[-\mu V_t ({\bf x})-\mu T \ln  p_t({\bf x}) \right] \frac{\partial p_t({\bf x})}{\partial t}\\
    &= \frac{1}{\mu T} \int d{\bf x} \ \Vert \bmnu_t ({\bf x}) \Vert ^2 p_t({\bf x}),
\end{align}
where we used Eq.~(\ref{fokker-planck}) and the normalization of the probability $(d/dt)[\int  d{\bf x} p_t ({\bf x}) ] =0$, and we assumed that $p_t({\bf x})$ vanishes at infinity.  The symbol $||\bmnu_t||^2 := \bmnu_t \cdot \bmnu_t$ indicates the square of $L^2$ norm. Thus, the entropy production rate $\sigma_t$ is given by the expected value of the square of the mean local velocity divided by the factor $\mu T$. The entropy production from time $t=0$ to time $t=\tau$ is defined as the time integral of the entropy production rate,
\begin{align}
    \Sigma & := \int^{\tau}_{0} dt \sigma_t.
\end{align}
We can easily check the non-negativity of the entropy production $\Sigma\geq 0$, which implies the second law of thermodynamics.

\subsection{Optimal transport theory and $L^2$-Wasserstein distance}
Next, we discuss the geometric measure of optimal transport called the $L^2$-Wasserstein distance~\cite{Villani2008Optiaml}.
We consider the cost function $c({\bf x},{\bf y})$ of transporting a single particle at the point ${\bf x} \in X$ to the point ${\bf y} \in X$. We first introduce the Monge-Kantrovich minimization problem~\cite{Kantrovich2006Transport}, which quantifies a difference between two probability distributions $p({\bf x})$ and $q({\bf y})$. The optimal transport cost for $c({\bf x},{\bf y})$ between $p({\bf x})$ and $q({\bf y})$ is defined as
\begin{align}
    C(p,q) := \min_{\Pi \in \mathcal{P}(p,q)} \int d{\bf x} d{\bf y} \ c({\bf x},{\bf y}) \Pi({\bf x},{\bf y}),
\end{align}
where the lower bound is taken over the entire set $\mathcal{P}(p,q)$ of joint probability distributions $\Pi({\bf x},{\bf y})$
on $X\times X$,
\begin{align}
    \mathcal{P}(p,q) := \{ \Pi |& p({\bf x}) =\int d{\bf y} \Pi ({\bf x},{\bf y}), \nonumber\\
    & q({\bf y}) =\int d{\bf x} \Pi ({\bf x},{\bf y}),  \Pi ({\bf x},{\bf y}) \geq 0  \},
\end{align}
where marginal distributions of $\Pi({\bf x},{\bf y})$ in the set $\mathcal{P}(p,q)$ are given by $p({\bf x})$ and $q({\bf y})$.
Therefore, the optimal transport cost gives a minimum value of the expected value of the cost $c({\bf x},{\bf y})$ for the joint distribution $\Pi({\bf x},{\bf y})$. We call the value of $\Pi$ that minimizes the expected value of the distance as the optimal transport plan $\Pi^{\ast}$, which is defined as
\begin{align}
    \Pi^{\ast} ({\bf x},{\bf y}) := \mathrm{argmin}_{\Pi \in \mathcal{P}(p,q)} \int d{\bf x} d{\bf y} \ c({\bf x},{\bf y}) \Pi({\bf x},{\bf y}) .
\end{align}
In general, the Monge-Kantrovich minimization problem is hard to be solved analytically. However, if we consider the $L^2$-norm as the optimal transport cost on the Euclidean space, the Monge-Kantrovich minimization problem can be solved with few restrictions~\cite{Villani2008Optiaml}. This optimal transport cost leads to the  $L^2$-Wasserstein distance which plays an important role in this paper.

The $L^2$-Wasserstein distance $\mathcal{W}(p,q)$ is introduced by the square root of the optimal transport cost for the cost function which is the square of the $L^2$-norm. Explicitly, the $L^2$-Wasserstein distance $\mathcal{W}(p,q)$ between $p$ and $q$ is defined as
\begin{align}
    \mathcal{W}(p,q)^2 &:= \min_{\Pi \in \mathcal{P}(p,q)} \int d{\bf x} d{\bf y} \ \Vert {\bf x}-{\bf y} \Vert^2 \Pi({\bf x},{\bf y}) \nonumber \\
    &= C(p, q),
\end{align}
where $C(p, q)$ is the optimal transport cost for the cost function $c({\bf x},{\bf y}) = \Vert {\bf x} - {\bf y} \Vert^2$.
The $L^2$-Wasserstein distance is well defined~\cite{Villani2008Optiaml} if two probability distributions $p$ and $q$ satisfy
\begin{align}
    \int d{\bf x} p({\bf x}) \Vert {\bf x} \Vert^2 <\infty , \int d{\bf y} q({\bf y}) \Vert {\bf y} \Vert^2 <\infty,
    \label{condition1}
\end{align}
which is only an assumption for two distributions $p$ and $q$ to define $L^2$-Wasserstein distance. We assume this condition Eq.~(\ref{condition1}) throughout the paper.

\begin{figure}
    \centering
        \includegraphics[width=\hsize]{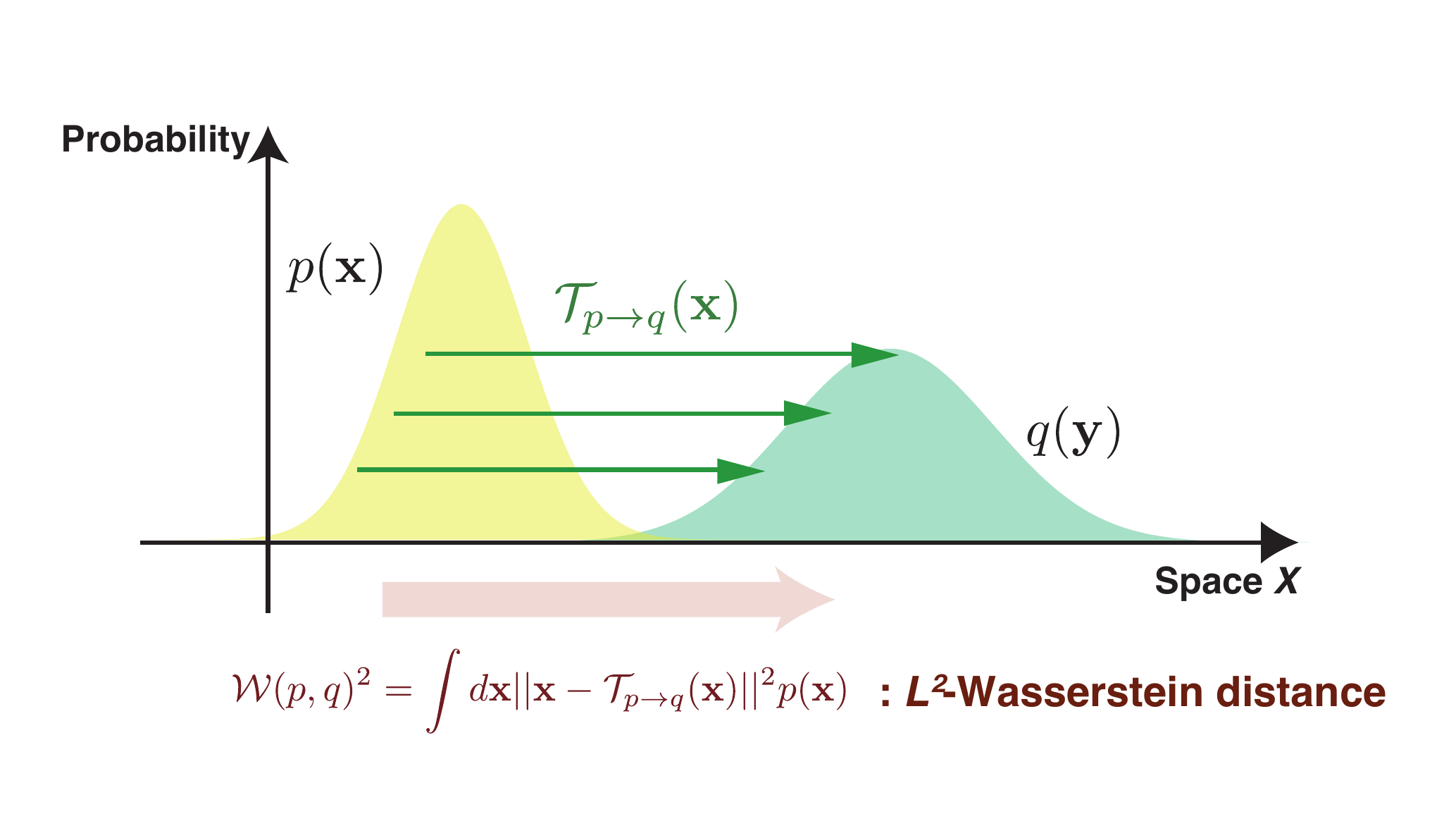}
        \caption{Schematic of the $L^2$-Wasserstein distance. We consider optimal transport from the probability distribution $p(\bf{x})$ to the probability distribution $q(\bf{y})$. The length of the green arrow shows the optimal transportation distance $|| {\bf x}-\mathcal{T}_{p \to q}({\bf x})||$, and the square of the $L^2$-Wasserstein distance is given by the expected value of the square of its optimal transportation distance.}
        \label{fig1}
\end{figure}
Furthermore, it is known that there exists a map $\mathcal{T}_{p \to q}({\bf x})$ such that $\Pi^{\ast}({\bf x},{\bf y}) = p({\bf x})\delta({\bf y}-\mathcal{T}_{p \to q}({\bf x}))$ for the $L^2$-Wasserstein distance $c({\bf x},{\bf y}) =\Vert {\bf x}- {\bf y} \Vert^2$ on the space $\mathbb{R}^d$, where $\delta({\bf x})$ is the delta function~\cite{Villani2008Optiaml}.
This map $\mathcal{T}_{p \to q}$ is called the optimal transport map from $p$ to $q$.
Using the fact that the marginal distributions of $\Pi^{\ast}({\bf x},{\bf y})$ are $p({\bf x})$ and $q({\bf y})$, we can obtain 
\begin{align}
\int d{\bf y} f({\bf y})q({\bf y}) &= \int d{\bf x} \int d{\bf y} f({\bf y})\Pi^{\ast}({\bf x},{\bf y})  \nonumber\\
&=\int d{\bf x} f(\mathcal{T}_{p \to q}({\bf x})) p({\bf x}) 
\label{optimaltransportint}
\end{align}
for any differential and measurable function $f({\bf x})$. If we consider the change of variables ${\bf y} = \mathcal{T}_{p \to q}({\bf x})$ and $d{\bf y} =d{\bf x}  \vert \det (\nabla \mathcal{T}_{p\rightarrow q}({\bf x})) \vert$, we obtain the Jacobian equation~\cite{Villani2008Optiaml}
\begin{align}
   p({\bf x})= q(\mathcal{T}_{p\rightarrow q} ({\bf x})) \vert \det (\nabla \mathcal{T}_{p\rightarrow q}({\bf x})) \vert,
   \label{jacobianeq}
\end{align} 
where $\vert \det (\nabla \mathcal{T}_{p \to q}({\bf x})) \vert$ denotes the determinant of the Jacobian matrix $\nabla \mathcal{T}_{p \to q}$ at ${\bf x}$. By using the optimal transport map, the $L^2$-Wasserstein distance is calculated as 
\begin{align}
    \mathcal{W}(p,q)^2 = \int d{\bf x} \ \Vert {\bf x}-\mathcal{T}_{p \to q}({\bf x}) \Vert^2 p({\bf x}).
    \label{wasseroptimal}
\end{align}
Thus, the $L^2$-Wasserstein distance can be regarded as the expected value of $|| {\bf x}-\mathcal{T}_{p \to q}({\bf x})||^2$ (see Fig.~\ref{fig1}).

We briefly introduce the Benamou-Brenier formula~\cite{Benamou2000Monge}, which is related to a relation between the entropy production and the $L^2$-Wasserstein distance in this paper. If dynamics of the probability $q_t({\bf x} )$ at time $t$ are driven by the continuity equation with the velocity field ${\bf v}_t({\bf x} )$,
\begin{align}
\frac{  \partial q_t({\bf x} )}{\partial t} = -\nabla \cdot ({\bf v}_t({\bf x} ) q_t({\bf x} )),
\end{align}
the $L^2$-Wasserstein distance gives the lower bound on the expected value of the square of the velocity field,
\begin{align}
\mathcal{W} (q_0, q_{\tau})^2 \leq \tau \int_0^{\tau} dt \int d{\bf x} ||{\bf v}_t({\bf x} ) ||^2 q_t({\bf x} ).
\end{align}
where we consider the time integral from time $t=0$ to time $t=\tau$. Because the velocity field of the Fokker-Planck equation is the mean local velocity, we obtain a relation between the entropy production rate and the $L^2$-Wasserstein distance as discussed in the next section.

\section{Entropy production and $L^2$-Wasserstein distance}
\subsection{Relation between Wasserstein distance and entropy production rate}
In this section, we discuss a relation between the $L^2$-Wasserstein distance and the entropy production rate, which is a main result in this paper. We set that dynamics of the probability distribution $p_t({\bf x})$ are described by the Fokker-Planck equation (\ref{fokker-planck}). 
We define the path length on the probability simplex measured by the $L^2$-Wasserstein distance from time $t=0$ to time $t=\tau$ as
\begin{align}
    \mathcal{L}_{\tau} := \lim_{\Delta t \rightarrow + 0} \sum_{k =0}^{n} 
    \mathcal{W}(p_{k \Delta t},p_{(k+1)\Delta t }),
    \label{pathlength}
\end{align}
where $n$ is a positive integer satisfying $n  \Delta t \leq \tau \leq (n+1)  \Delta t$. The entropy production rate is bounded by
\begin{align}
\sigma_t \geq   \frac{1}{\mu T}\left(\frac{d \mathcal{L}_t}{ d t } \right)^2 , 
\label{syutyouineq} 
\end{align}
which is a main result in this paper. This main result is consistent with the optimal transport theory for an infinitesimal time transition in Ref.~\cite{Villani2000generalization}. This equation gives a relation between the $L^2$-Wasserstein distance and the entropy production rate for the Fokker-Planck equation. In terms of the $L^2$-Wasserstein distance, the quantity $(d \mathcal{L}_t/ d t  )^2$ is given by
\begin{align}
\left(\frac{d \mathcal{L}_t}{ d t } \right)^2 &= \lim_{\Delta t \rightarrow +0} \frac{ \mathcal{W}(p_{t + \Delta t} , p_t)^2}{ \Delta t^2}.
\end{align}
Thus, this inequality can be regarded as the Benamou-Brenier formula~\cite{Benamou2000Monge} for the short time $\tau=\Delta t$,
\begin{align}
\mathcal{W}(p_{t + \Delta t} , p_t)^2 \leq \Delta t \int_0^{\Delta t} dt \int d{\bf x} \ \Vert \bmnu_t ({\bf x}) \Vert ^2 p_t({\bf x}) + O(\Delta t^3).
\end{align}

We next discuss the situation that the equality in Eq.~(\ref{syutyouineq}) holds. We introduce a non-negative term $\sigma_t^{\rm rot}$ defined as
\begin{align}
 \sigma_t^{\rm rot}&= \sigma_t - \frac{1}{\mu T}\left(\frac{d \mathcal{L}_t}{ d t } \right)^2 \geq 0,
\label{syutyou2} 
\end{align}
and discuss the situation $\sigma_t^{\rm rot}=0$. We consider the Taylor expansion of the optimal transport map $\mathcal{T}_{p_t\rightarrow p_{t+\Delta t}}({\bf x}) $ up to the order $\Delta t$,
\begin{align}
    &\mathcal{T}_{p_t\rightarrow p_{t+\Delta t}}({\bf x}) = {\bf x} + \bm{a}_1({\bf x})\Delta t + \mathcal{O}(\Delta t^2),   \label{taylorexp1}
\end{align}
where $\bm{a}_1({\bf x})$ is the first order of the Taylor coefficient. From Eq.~(\ref{wasseroptimal}), we obtain an expression of $ \left(d\mathcal{L}_t/ dt\right)^2$,
\begin{align}
\left(\frac{d \mathcal{L}_t}{ d t } \right)^2 &= \int d{\bf x} \ \Vert \bm{a}_1({\bf x}) \Vert ^2 p_t({\bf x}) .
\label{expression1}
\end{align}
Thus, if the mean local velocity gives an optimal transport map, that is $\bmnu_t ({\bf x})= \bm{a}_1({\bf x})$, the equality holds and $ \sigma_t^{\rm rot} =0$. 

Next, we consider the difference between $\bm{a}_1({\bf x})$ and $\bmnu_t ({\bf x})$. By substituting $(p_{t},p_{t+\Delta t} )$ into $(p,q)$, the Jacobian equation in Eq.~(\ref{jacobianeq}) is given by
\begin{align}
    p_t({\bf x})= p_{t+\Delta t}(\mathcal{T}_{p_t\rightarrow p_{t+\Delta t}} ({\bf x})) \vert \det (\nabla \mathcal{T}_{p_t\rightarrow p_{t+\Delta t}}({\bf x})) \vert.
    \label{seishitu}
    \end{align}
We calculate the Taylor expansions of the determinant up to the order $\Delta t$ as follows
\begin{align}
 \vert \det (\nabla \mathcal{T}_{p_t\rightarrow p_{t+\Delta t}}({\bf x})) \vert = 1 + \nabla \cdot \bm{a}_1({\bf x})\Delta t + \mathcal{O}(\Delta t^2).
    \label{taylorexp2}
\end{align}
From the Fokker-Planck equation (\ref{fokker-planck}), we also obtain 
\begin{align}
&p_{t+\Delta t}({\bf x}) = p_t({\bf x}) - \nabla \cdot(\bm{\nu}_t({\bf x})p_t({\bf x}))\Delta t + \mathcal{O}(\Delta t^2),
    \label{taylorexp3}
\end{align}
which is the discretized version of the Fokker-Planck equation for the short time $\Delta t$.
By inserting Eqs.~(\ref{taylorexp1}), (\ref{taylorexp2}) and (\ref{taylorexp3}) into Eq.~(\ref{seishitu}), we obtain
\begin{align}
0=  \nabla \cdot [(\bm{a}_1({\bf x})-\bm{\nu}_t({\bf x}))p_t({\bf x})] \Delta t \notag + \mathcal{O}(\Delta t^2).
\end{align}
By considering the first-order terms of $\Delta t$, we obtain
\begin{align}
\nabla \cdot [(\bm{a}_1({\bf x})-\bm{\nu}_t({\bf x}))p_t({\bf x})]=0.
\label{conditiona}
\end{align}
In case of $d=3$, this equation implies the existence of a vector potential  $\boldsymbol{A}_t ({\bf x})$ because of Helmholtz's decomposition  such as
\begin{align}
\bm{a}_1({\bf x})p_t({\bf x}) = \bm{\nu}_t({\bf x})p_t({\bf x}) + \nabla \times \boldsymbol{A}_t ({\bf x}).
\label{helmholtz}
\end{align}
 Thus, this vector potential $\boldsymbol{A}_t$ quantifies a difference between optimal transport plan and the time evolution of the Fokker-Planck equation from time $t$ to time $t+\Delta t$. In the general case of $d \neq 3$ or the case of non-Euclidean space, we may consider the Helmholtz-Hodge decomposition to obtain an expression of $\bm{a}_1({\bf x})p_t({\bf x}) - \bm{\nu}_t({\bf x})p_t({\bf x})$.

To find the expression of $\sigma^{\rm rot}_t$, we use the formula for the time derivative of the $L^2$-Wasserstein distance~\cite{Villani2008Optiaml}. The following formula
    \begin{align}
        &\left. \frac{d}{ds}\left(\frac{\mathcal{W}(p, p_{t+s})^2}{2} \right) \right|_{s=0}\! =\! - \!\int d{\bf x}  ({\bf x}\!-\!\mathcal{T}_t({\bf x}) )\!\cdot \!\bmnu_t (\mathcal{T}_t({\bf x}))p({\bf x})  \label{theorem}
    \end{align}
holds for any probability distribution $p({\bf x})$, where we used the notation $\mathcal{T}_t=\mathcal{T}_{p \to p_t}$. The proof of this formula (\ref{theorem}) is shown in Appendix A. By applying the Taylor expansion Eq.~(\ref{taylorexp1}) to the formula Eq.~(\ref{theorem}) for $(p, p_{t+s})=(p_t, p_{t+\Delta t +s})$, we obtain the following equation,
\begin{align}
    &\left. \frac{d}{ds}\left(\frac{\mathcal{W}(p_t, p_{t+\Delta t+s})^2}{2} \right) \right|_{s=0}\notag\\
     =&\Delta t \int d{\bf x}  [\bm{a}_1({\bf x})\cdot \bmnu_t (\mathcal{T}_{p_t\rightarrow p_{t+\Delta t}}({\bf x}))]p_t({\bf x}) \notag\\
     =& \Delta t \!\int \! d{\bf x} [\bm{a}_1(\mathcal{T}_{p_t\rightarrow p_{t+\Delta t}}({\bf x}))\!\cdot \!\bmnu_t (\mathcal{T}_{p_t\rightarrow p_{t+\Delta t}}({\bf x}))]p_t({\bf x}) \!+\!\mathcal{O}(\Delta t^2) \notag\\
     =& \Delta t \int d{\bf y} [\bm{a}_1({\bf y})\cdot \bmnu_t ({\bf y})] p_{t+\Delta t} ({\bf y})  +\mathcal{O}(\Delta t^2)\notag\\
    =& \Delta t \int d{\bf x} [\bm{a}_1({\bf x})\cdot \bmnu_t ({\bf x}) ]p_{t} ({\bf x}) +\mathcal{O}(\Delta t^2), \label{mousugu}
\end{align}
where we used Eq.~(\ref{optimaltransportint}). From the definition of the path length Eq.~(\ref{pathlength}), we obtain
\begin{align}
\mathcal{W}(p_{t+s},p_{t}) =  \frac{d\mathcal{L}_t}{dt} s + \mathcal{O} (s^2),
\end{align}
for small $s$.
Therefore, we also obtain 
\begin{align}
    &\left. \frac{d}{ds}\left(\frac{\mathcal{W}(p_t, p_{t+\Delta t+s})^2}{2} \right) \right|_{s=0}\notag\\
    &= \lim_{s\rightarrow + 0} \frac{\mathcal{W}(p_{t+\Delta t+s},p_t) -\mathcal{W}(p_{t+\Delta t},p_t)}{s}  \mathcal{W}(p_{t+\Delta t},p_t) \notag \\
    &= \frac{d\mathcal{L}_{t+\Delta t} }{dt} \frac{d\mathcal{L}_t}{dt} \Delta t + \mathcal{O}(\Delta t^2) \notag \\
    &= \left( \frac{d\mathcal{L}_t}{dt}\right)^2\Delta t + \mathcal{O}(\Delta t^2).
    \label{pathderivativesquare}
\end{align}
By comparing Eq.~(\ref{pathderivativesquare}) with Eq. (\ref{mousugu}), we obtain another expression of $ \left(d\mathcal{L}_t/ dt\right)^2$,
\begin{align}
   \left( \frac{d\mathcal{L}_t}{dt}\right)^2 =& \int d{\bf x}[ \bm{a}_1({\bf x})\cdot \bmnu_t ({\bf x}) ]p_{t} ({\bf x}) .
   \label{expression2}
\end{align}
We also obtain expressions of $\sigma^{\rm rot}_t$,
\begin{align}
   \sigma^{\rm rot}_t =& \frac{1}{\mu T}\int d{\bf x}[ (\bmnu_t ({\bf x})  - \bm{a}_1({\bf x}) )\cdot \bmnu_t ({\bf x}) ]p_{t} ({\bf x}) , \\
   &= \frac{1}{\mu T} \int d{\bf x} ||\bmnu_t ({\bf x}) - \bm{a}_1({\bf x})  ||^2 p_{t} ({\bf x}),
\end{align}
where we compared Eq.~(\ref{expression1}) with Eq.~(\ref{expression2}). Thus, $\sigma^{\rm rot}_t$ is non-negative, and zero if $|| \bm{a}_1({\bf x}) - \bmnu_t ({\bf x})  ||= 0$. This value $\sigma^{\rm rot}_t$ quantifies the amount of a difference between the velocity field of optimal transport and the mean local velocity.  In case of $d=3$,  $\sigma^{\rm rot}_t$ is calculated as
\begin{align}
    \sigma^{\rm rot}_t &= - \frac{1}{\mu T} \int d{\bf x} [\nabla \times \boldsymbol{A}_t ({\bf x}) ]\cdot \bm{\nu}_t ({\bf x}) \\
    &= \frac{1}{\mu T} \int d{\bf x} \frac{|| \nabla \times \boldsymbol{A}_t ({\bf x}) ||^2}{ p_{t} ({\bf x})} \geq 0,
\end{align}
which quantifies the amount of the rotation because $\sigma^{\rm rot}_t$ is proportional to the mean value of the square of the rotation $|| \nabla \times \boldsymbol{A}_t ({\bf x}) ||/p_t({\bf x})$.
Thus, $\sigma^{\rm rot}_t$ is non-negative, and zero if the rotation vanishes $|| \nabla \times \boldsymbol{A}_t ({\bf x}) ||= 0$.

We discuss when $\sigma^{\rm rot}_t$ vanishes. If the mean local velocity $\bm{\nu}_t ({\bf x})$  is given by $\bm{\nu}_t ({\bf x}) = - \nabla \Phi_t$ with the potential $\Phi_t = \mu (V_t(\bm{x}) + T \ln p_t(\bm{x}))$ as we assumed in Eq.~(\ref{fokker-planckassume}),  the quantity $\sigma^{\rm rot}_t$ is given by 
\begin{align}
   \sigma^{\rm rot}_t &= -\frac{1}{\mu T}\int d{\bf x}[ (\bmnu_t ({\bf x})  - \bm{a}_1({\bf x}) )p_{t} ({\bf x})] \cdot  \nabla \Phi_t \notag\\
   &=-\frac{1}{\mu T}\int d{\bf x} \nabla \cdot   [ (\bmnu_t ({\bf x})  - \bm{a}_1({\bf x}) )p_{t} ({\bf x})  \Phi_t] \notag\\
   &=-\frac{1}{\mu T} \int d\boldsymbol{S} \cdot [ (\bmnu_t ({\bf x})  - \bm{a}_1({\bf x}) )p_{t} ({\bf x})  \Phi_t]
\end{align}
where we used Eq.~(\ref{conditiona}) and $\int d\boldsymbol{S}$ denotes the surface integral. If the quantity $\Vert  [ (\bmnu_t ({\bf x})  - \bm{a}_1({\bf x}) )p_{t} \Phi_t \Vert $ vanishes at infinity, the quantity $\sigma^{\rm rot}_t$ becomes zero. The assumption that the probability $p_{t}$ vanishes at infinity is physically natural. Therefore, $\sigma^{\rm rot}_t$ vanishes in a physically natural situation and we obtain
\begin{align}
\sigma_t =  \frac{1}{\mu T}\left(\frac{d \mathcal{L}_t}{ d t } \right)^2.
\label{syutyou}
\end{align}
The above condition $\bm{\nu}_t ({\bf x}) = - \nabla \Phi_t$ is based on the assumption that the force $- \nabla V_t(\bm{x})$ is given by the gradient of the potential $V_t(\bm{x})$. In the section~V B, we analytically show $\sigma^{\rm rot}_t=0$ for the 1D-Brownian particle trapped in the harmonic potential. If the force is given by the non-poteintial force and the mean local velocity $\bm{\nu}_t ({\bf x})$ is not given by the potential, the term $\sigma^{\rm rot}_t$ might not be zero. The non-poteintial force is needed to achieve nonequilibrium steady state, and the steady flow and the steady force should be cyclic because of the Schnakenberg network theory~\cite{Schnakenberg1976network}. The quantity $\sigma^{\rm rot}_t$ might play an important role in the steady state thermodynamics with the non-potential force~\cite{Hatano2001steady}.

\subsection{Geometric lower bounds on entropy production and thermodynamic speed limits}
\begin{figure}
\centering
        \includegraphics[width=\hsize]{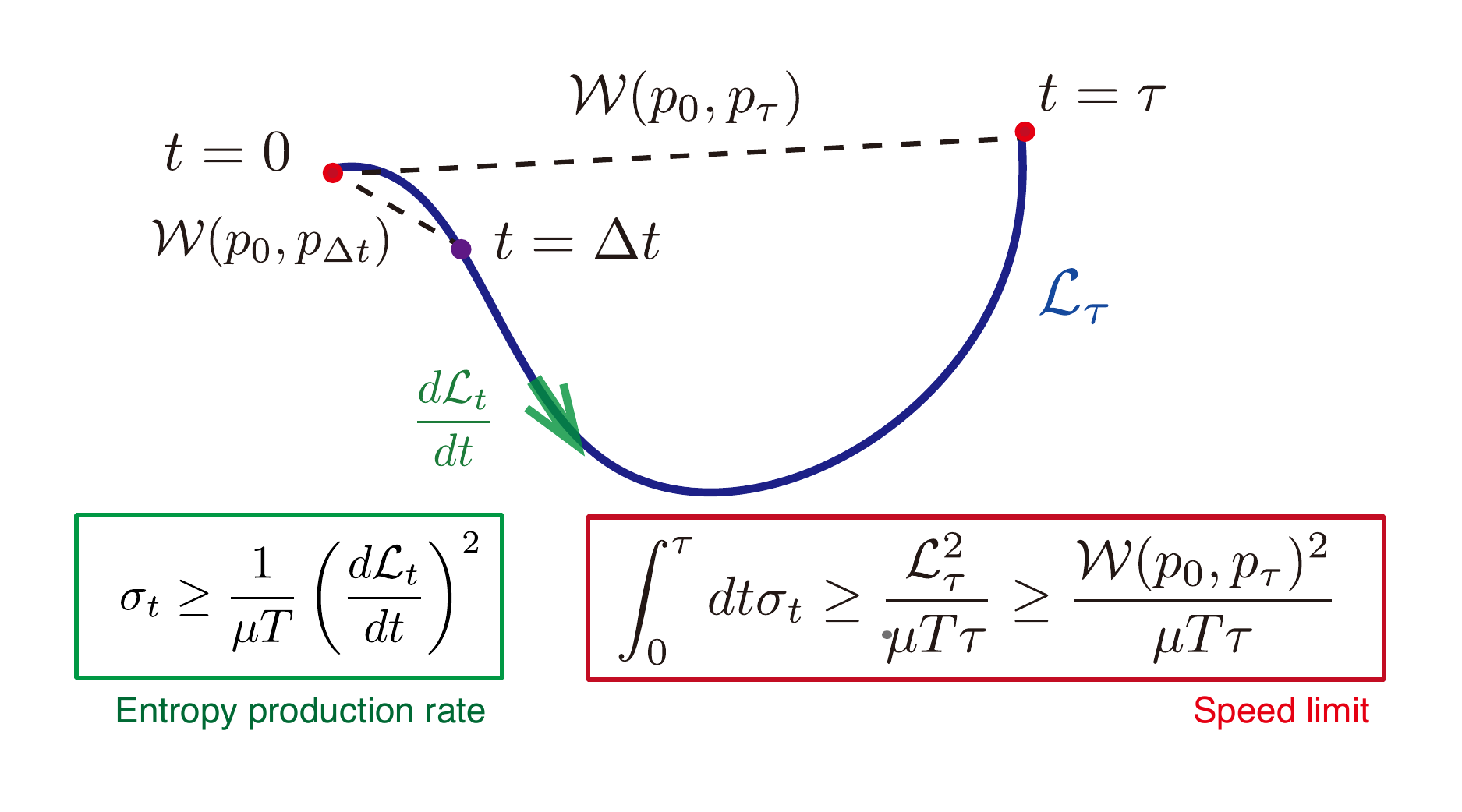}
        \caption{Schematic of the entropy production and the $L^2$-Wasserstein distance. The lower bound on the entropy production is obtained from geometry of the $L^2$-Wasserstein distance. The entropy production $\Sigma = \int_0^{\tau} dt \sigma_t$ is tightly bounded by the length measured by the $L^2$-Wasserstein distance $\mathcal{L}_{\tau}$. The $L^2$-Wasserstein distance itself $\mathcal{W} (p_0 ,p_{\tau})$ provides a looser bound on the entropy production.
        These inequalities are generalizations of thermodynamic speed limits.}
        \label{fig2}
\end{figure}

We here discuss a lower bound on the entropy production $\Sigma := \int dt \sigma_t$ based on Eq. (\ref{syutyouineq}). By using Eq. (\ref{syutyouineq}), the entropy production from time $t= 0$ to time $t= \tau$ is bounded by 
\begin{align}
    \Sigma & = \int^{\tau}_{0} dt \sigma_t  \notag \\
    &\geq \frac{1}{\mu T} \int^{\tau}_{0}dt \left( \frac{d \mathcal{L}_t}{dt}\right)^2.
\end{align}
In differential geometry, the quantity $\mathcal{C}= (1/2)\int^{\tau}_{0}dt \left(d \mathcal{L}_t/ dt\right)^2$ is called the action, and Eq. (\ref{syutyouineq}) implies that the entropy production for the Fokker-Planck equation is bounded by the action measured by the path length of the Wasserstein $L^2$ distance, 
\begin{align}
\Sigma \geq \frac{2 \mathcal{C}}{\mu T}.
\end{align}
If $\sigma^{\rm rot} =0$, the entropy production is proportional to the action measured by the path length of the Wasserstein $L^2$ distance $\Sigma  = 2 \mathcal{C}/(\mu T)$.
Here, we consider the following Cauchy-Schwarz inequality 
\begin{align}
    2 \tau \mathcal{C} &=\left(\int^{\tau}_{0}dt\right)\left(\int^{\tau}_{0} dt \left(\frac{d \mathcal{L}_t}{dt}\right)^2\right) \notag \\
    &\geq \left(\int^{\tau}_{0} dt \frac{d \mathcal{L}}{dt}\right)^2 \notag \\
    &= \mathcal{L}_{\tau}^2,
    \label{cauchyschwarz}
\end{align}
which gives a lower bound on the action. In information geometry, this inequality has been considered~\cite{crooks2007measuring} as a trade-off relation between time $\tau$ and the action $\mathcal{C}$. By considering $\left(d \mathcal{L}_t/ dt\right)^2$ as the Fisher information of time, several variants of thermodynamic speed limits can be derived from this inequality for the Markov jump process~\cite{ito2018stochastic1}, the Fokker-Planck equation~\cite{ito2020stochastic} and the rate equation~\cite{yoshimura2020information} in stochastic thermodynamics of information geometry. 
In the same way, we obtain a lower bound on the entropy production by considering the action measured by the $L^2$-Wasserstein distance  (see also Fig.~\ref{fig2}),
\begin{align}
    \Sigma \geq \frac{\mathcal{L}_{\tau}^2}{\mu T \tau },
    \label{newInequality}
\end{align}
which is also a main result in this paper. 
Because this inequality implies a trade-off relation between time and the entropy production, this result can also be regarded as a generalization of thermodynamic speed limits.
Since we use the Cauchy-Schwarz inequality, the
equality can be achieved when the probability distribution moves with a constant velocity on the $L^2$-Wasserstein distance space, 
that is, when it satisfies the following equation
\begin{align}
    \frac{d \mathcal{L}_t}{dt} = \frac{\mathcal{L}_{\tau} }{\tau},
\end{align}
for any $0 \leq t \leq \tau$.

Using the fact that the $L^2$-Wasserstein distance satisfies the triangle inequality for probabilities $p$, $q$ and $r$~\cite{Villani2008Optiaml}
\begin{align}
   \mathcal{W}(p,r) \leq \mathcal{W}(p,q) + \mathcal{W}(q,r),
    \label{triangle}
\end{align}
we obtain the following inequality,
\begin{align}
    \mathcal{L}_{\tau}  \geq \mathcal{W}(p_{0},p_{\tau}) .
\end{align}
from the definition of $\mathcal{L}_{\tau}$. 
Using Eq.~(\ref{newInequality}) and the above inequality, we can obtain the previously known inequality in Refs.~\cite{Aurell2012Refined,  Dechant2019Wasserstein},
\begin{align}
    \Sigma \geq \frac{\mathcal{W}(p_{0},p_{\tau})^2}{\mu T \tau}, \label{oldInequality}
\end{align}
which is equivalent to the Benamou-Brenier formula~\cite{Benamou2000Monge} because the entropy production rate is given by the expected value of the square of the velocity field $\boldsymbol{ \nu}_t (\bf{x}) $.
Considering the above derivation, 
the condition for the equality to hold is when the probability distribution changes 
at a constant speed on a straight line as measured by the $L^2$-Wasserstein distance,
\begin{align}
    \mathcal{L}_{\tau}  &= \mathcal{W}(p_{\tau} , p_{0}), \\
    \frac{d \mathcal{L}_t}{dt} &= \frac{ \mathcal{W}(p_{\tau} , p_{0})}{\tau} .
    \label{minimization}
\end{align}
In this case, the entropy production is minimized with constraints $p_0$ and $p_{\tau}$. Moreover, when the initial distribution $p_{0}$, the final distribution $p_{\tau}$, 
and the time interval $\tau$ are specified, the protocol to achieve this equality can be numerically obtained by the algorithm of the fluid mechanics~\cite{Benamou2000Monge}.  In other words, by using this algorithm, we can construct an efficient heat engine for small systems with the minimum entropy production.

Similarly, we obtain another lower bound by applying the Cauchy-Schwartz inequality Eq.~(\ref{cauchyschwarz}) and the triangle inequality Eq.~(\ref{triangle}). Let us consider the time interval $t_i= \tau (i/N)$. Because the entropy production is given by
\begin{align}
    \Sigma & \geq \sum_{i=0}^{N-1} \frac{1}{\mu T} \int^{t_{i+1}}_{t_i}dt \left( \frac{d \mathcal{L}_t}{dt}\right)^2,
\end{align}
another lower bound on the entropy production can be obtained in a similar way as follows
\begin{align}
  \Sigma &\geq \sum_{i=0}^{N-1} \hat{\Sigma}(t_i;t_{i+1}),
  \label{estimationbound}
  \end{align}
 where $\hat{\Sigma}(t;s) $ is the lower bound on the entropy production by the Benamou-Brenier formula from time $t$ to time $s$,
  \begin{align}
  \hat{\Sigma}(t;s) &= \frac{\mathcal{W}(p_{t},p_{s})^2}{\mu T (s-t)}.
\end{align}
Moreover, in case of $\sigma_t^{\rm rot} =0$, we obtain
\begin{align}
  \Sigma &= \lim_{N \to \infty} \sum_{i=0}^{N-1} \hat{\Sigma}(t_i;t_{i+1}), 
  \label{estimationentropyproduction}
\end{align}
because the change from $p_{t_i}$ to $p_{t_{i+1}}$ is at a constant rate on a straight line as measured by the $L^2$-Wasserstein distance in the limit $t_{i+1} - t_i = \tau/N \to 0$. Remarkably, a calculation of $\hat{\Sigma}(t_i;t_{i+1})$ does not require information of the joint probability distribution at time $t_i$ and $t_{i+1}$, while the experimental estimation of the entropy production based on the fluctuation theorem needs information of the joint probability distribution~\cite{Ciliberto2017experiments}. It is relatively difficult to estimate the joint probability in an experiment with a small number of samples, compared to two probabilities. 
This fact might be useful to estimate the entropy production in an experiment by using Eq. (\ref{estimationentropyproduction}). This estimation of the entropy production by using Eq. (\ref{estimationentropyproduction}) is similar to the estimation of the entropy production based on thermodynamic trade-off relations such as thermodynamic uncertainty relations~\cite{Horowitz2019quantifying,Manikandan2020inferring,otsubo2020estimation,Vu2020entropy}. 

\section{subsystem and information thermodynamics}

\subsection{ Stochastic thermodynamics for subsystem}
Stochastic thermodynamics for a subsystem has been discussed in terms of information thermodynamics, which explains a paradox of the Maxwell's demon~\cite{parrondo2015thermodynamics}. In information thermodynamics, we consider a relation between the partial entropy production and information flow for the 2D Fokker-Planck equation (\ref{subsystem}) or the 2D Langevin equations~\cite{allahverdyan2009thermodynamic, horowitz2014secondlaw,ito2015maxwell}.  In this section, we discuss a relationship between the partial entropy production and the $L^2$-Wasserstein distance for a subsystem.
 
We start with two-dimensional systems $X$ and $Y$. Stochastic dynamics of two positions $x \in X (= \mathbb{R})$ and $y \in Y (=\mathbb{R})$ are driven by the following Fokker-Planck equation
\begin{align}
    \frac{\partial p_t(x,y)}{\partial t}\! =& \!-\!\frac{\partial}{\partial x}  [\nu^X_t \!(x,y)p_t (x,y)] \!-\!\frac{\partial}{\partial y}  [\nu^Y_t \!(x,y)p_t (x,y)], \notag\\
   \nu^X_t (x,y) :=&-\mu \frac{\partial}{\partial x} [V_t (x,y)  +T \ln p_t(x,y)], \notag\\
   \nu^Y_t (x,y) :=&-\mu\frac{\partial}{\partial y} [V_t (x,y)  +T \ln p_t(x,y)]. \label{subsystem}
\end{align}

We first consider the situation that the position $y$ is the hidden degree of freedom and we can only observe the position $x$. 
Thus, we can only measure the marginal distribution of $X$ defined as
\begin{align}
  p^X_t(x) = \int dy \ p_t(x,y). 
\end{align}
The time evolution of the marginal distribution is given by
\begin{align}
    \frac{\partial p^X_t(x)}{\partial t} &= - \frac{\partial}{\partial x}  \left ( \bar{\nu}^X_t(x)  p^X_t(x) \right ), 
    \end{align}
    \begin{align}
\bar{\nu}^X_t(x)
    &= \frac{\int dy  \nu^X_t(x,y)p_t(x,y)}{p^X_t(x)} \notag \\
    &= \int dy \nu^X_t(x,y)p^{Y|X}_t(y|x),
\end{align}
where $\bar{\nu}_t^X(x)$ is the marginal mean local velocity of $X$ and $p^{Y|X}_t(y|x) := p_t(x,y)/ p^X_t(x)$ is the conditional probability of $Y$ under the condition of $X$. If we want to measure the entropy production rate for this system, we only obtain the apparent entropy production rate of $X$,
\begin{align}
    {\bar{\sigma}}^X_t = \frac{1}{\mu T}\int dx [\bar{\nu}^X_t(x)]^2 p^X_t(x), 
\end{align}
which is different from the partial entropy production rate of $X$,
\begin{align}
    {\sigma}^X_t = \frac{1}{\mu T}\int dx \int dy [\nu^X_t(x,y)]^2 p_t(x, y).
\end{align}
From the Cauchy-Schwarz inequality, we obtain the inequality
\begin{align}
    &{\sigma}^X_t - {\bar{\sigma}}^X_t \notag\\
    =&\frac{1}{\mu T}  \int dx \frac{\left(\int dy  [\nu^X_t(x,y)]^2 p_t(x, y) \right) \left(\int dy p_t(x, y) \right)}{p^X_t(x)}\notag \\
    &-\frac{1}{\mu T}  \int dx \frac{\left( \int dy \nu^X_t(x,y)p_t(x,y) \right)^2}{p^X_t(x)} \notag\\
    \geq& 0. \label{appaent}
\end{align}
Thus, the apparent entropy production rate ${\bar{\sigma}}^X_t$ is always smaller than the partial entropy production rate ${\sigma}^X_t$. The apparent entropy production rate is equivalent to the partial entropy production when $\nu^X_t(x,y) =\bar{\nu}^X_t(x)$. This condition implies that the potential force $-\partial V_t (x,y)/ \partial x$ does not depend on $y$, and the systems $X$ and $Y$ are statistically independent $p_t(x,y)=p^X_t(x)p^Y_t(y)$ with $p^Y_t(y):= \int dx \ p_t(x,y)$.

If we define the path length of $X$ from time $t=0$ to time $t=\tau$ as
\begin{align}
    \mathcal{L}^X_{\tau} := \lim_{\Delta t \rightarrow + 0} \sum_{k =0}^{n} 
    \mathcal{W}(p^X_{k \Delta t},p^X_{(k+1)\Delta t }),
\end{align}
our result for the path length of $X$ gives the apparent entropy production rate of $X$,
\begin{align}
 {\bar{\sigma}}^X_t  \geq  \frac{1}{\mu T}\left(\frac{d \mathcal{L}^X_t}{ d t } \right)^2 =\lim_{\Delta t\to 0} \frac{\mathcal{W} (p^X_{t+\Delta t}, p^X_t)^2}{\mu T \Delta t^2},
 \label{syuthou2}
\end{align}
where $n$ is a positive integer satisfying $n \Delta t \geq \tau \geq (n+1) \Delta t$.
We also obtain a lower bound on the apparent entropy production rate of $X$ as follows,
\begin{align}
 {\bar{\Sigma}}^X &:= \int_0^{\tau} dt {\bar{\sigma}}^X_t \\
 &\geq   \frac{(\mathcal{L}^X_{\tau})^2}{ \tau \mu T} \label{ineqX1}\\
  &\geq   \frac{ \mathcal{W}(p^X_{0},p^X_{\tau})^2}{ \tau \mu T}. \label{ineqX2}
\end{align}
Compound inequalities of Eqs. (\ref{appaent}), (\ref{ineqX1}) and (\ref{ineqX2}) can be 446
regarded as an information-thermodynamic speed limit for the 447
subsystem.
We remark that $\bar{\nu}^X_t(x)$ is given by the potential $\bar{\nu}^X_t(x) = - \partial_x \Phi_t$ when $\bar{\nu}^X_t(x)=\nu^X_t(x,y)$. In this case, we may obtain ${\sigma}^X_t = {\bar{\sigma}}^X_t =(d \mathcal{L}^X_t/ d t )^2 /(\mu T)$ because of the same reason in case of $\sigma^{\rm tot}_t=0$ for the total system. We also remark that $\bar{\nu}^X_t(x)$ is also given by $\bar{\nu}^X_t(x) = - \partial_x \bar{\Phi}^X_t$ with the potential $\bar{\Phi}^X_t =\int dy p^Y (y)\Phi_t$ if the systems $X$ and $Y$ are the statistically independent $p_t(x,y)=p^X_t(x)p^Y_t(y)$.

Now, we discuss a relationship between the subsystem $X$ and the subsystem $Y$. We introduce the marginal mean local velocity of $Y$, the apparent entropy production rate of $Y$  and the partial entropy production rate of $Y$ as follows,
\begin{align}
  \frac{\partial p^Y_t(y)}{\partial t} &= -\frac{\partial}{\partial y}  \left ( \bar{\nu}^Y_t(y)  p^Y_t(y) \right ), \\
  \bar{\nu}^Y_t(y) &= \frac{\int dx \  \nu^Y_t(x,y)p_t(x,y)}{p^Y_t(y)}, \\
  {\bar{\sigma}}^Y_t &= \frac{1}{\mu T}\int dy [\bar{\nu}^Y_t(y)]^2 p^{Y}_t(y), \\
  {\sigma}^Y_t &= \frac{1}{\mu T}\int dx \int dy [\nu^Y_t(x,y)]^2 p_t(x,y).
\end{align}
We also obtain a lower bound on the apparent entropy production rate of $Y$ as follows,
\begin{align}
 {\bar{\Sigma}}^Y &:= \int_0^{\tau} dt {\bar{\sigma}}^Y_t \\
 &\geq   \frac{(\mathcal{L}^Y_{\tau})^2}{ \tau \mu T} \label{ineqY1}\\
 &\geq \frac{ \mathcal{W}(p^Y_{0},p^Y_{\tau})^2}{ \tau \mu T},  \label{ineqY2}\\
\mathcal{L}^Y_{\tau} &:= \lim_{\Delta t \rightarrow + 0} \sum_{k =0}^{n} \mathcal{W}(p^Y_{k \Delta t},p^Y_{(k+1)\Delta t }),
\end{align}
where $n$ is a positive integer satisfying $n \Delta t \leq \tau \leq (n+1) \Delta t$.
By definition, the entropy production rate is given by the sum of the partial entropy production rates,
\begin{align}
 \sigma_t = \sigma^X_t + \sigma^Y_t. 
\end{align}
Because $\sigma^X_t \geq {\bar{\sigma}}^X_t$ and $\sigma^Y_t \geq {\bar{\sigma}}^Y_t$, the inequality
\begin{align}
    \sigma_t -{\bar{\sigma}}^X_t -{\bar{\sigma}}^Y_t\geq 0,
    \label{entropyhierachy}
\end{align}
is satisfied. Thus, we obtain lower bounds on the entropy production from Eqs. (\ref{ineqX1}), (\ref{ineqX2}), (\ref{ineqY1}), (\ref{ineqY2}) and (\ref{entropyhierachy}),
\begin{align}
    \Sigma &\geq  \frac{(\mathcal{L}^X_{\tau})^2 +(\mathcal{L}^Y_{\tau})^2  }{ \tau \mu T } \\
    & \geq \frac{\mathcal{W} (p_0^X, p_{\tau}^X)^2 +\mathcal{W} (p_0^Y, p_{\tau}^Y)^2 }{ \tau \mu T }.
\end{align}
From the non-negativity of $(\mathcal{L}^Y_{\tau})^2$ and $\mathcal{W} (p_0^Y, p_{\tau}^Y)$, we also obtain 
\begin{align}
    \Sigma &\geq  \frac{(\mathcal{L}^X_{\tau})^2}{ \tau \mu T } \\
    & \geq \frac{\mathcal{W} (p_0^X, p_{\tau}^X)^2 }{ \tau \mu T },
\end{align}
as looser bounds.
This result implies that the entropy production of the total system is generally bounded by geometry of the $L^2$-Wasserstein distance for two subsystems $X$ and $Y$.

\subsection{Information thermodynamics}
We next discuss an interpretation of the above result based on information thermodynamics. In information thermodynamics, we consider the following decomposition of the partial entropy production ${\sigma}^X_t$ into informational term $-\dot{\mathcal{I}}^X$ and thermodynamic terms $\sigma^X_{{\rm bath};t}+\sigma^X_{{\rm sys};t}$. 
The partial entropy production rates of $X$ and $Y$ for Eq.~(\ref{subsystem}) are calculated as
\begin{align}
    {\sigma}^X_t &= \sigma^X_{{\rm bath};t}+\sigma^X_{{\rm sys};t}-\dot{\mathcal{I}}^X , \label{decomp} \\
    {\sigma}^Y_t &= \sigma^Y_{{\rm bath};t}+\sigma^Y_{{\rm sys};t}-\dot{\mathcal{I}}^Y ,\label{decomp2} \\
    \sigma_{{\rm bath}; t}^X \!&=  \frac{1}{ T}\int dx \int dy \left[-\frac{\partial  V_t(x,y)}{\partial x}  \right]\nu^X_t(x,y) p_t(x, y),\\
    \sigma_{{\rm bath}; t}^Y \!&=  \frac{1}{ T}\int dx \int dy \left[-\frac{\partial  V_t(x,y)}{\partial y}  \right]\nu^Y_t(x,y) p_t(x, y), \\
    \sigma_{{\rm sys}; t}^X &=  \int dx \int dy \left[-\frac{\partial  \ln p^X_t(x)}{\partial x}  \right]\nu^X_t(x,y) p_t(x, y), \\
    \sigma_{{\rm sys}; t}^Y &=  \int dx \int dy \left[-\frac{\partial  \ln p^Y_t(y)}{\partial y}  \right] \nu^Y_t(x,y) p_t(x, y), \\
    \dot{\mathcal{I}}^X \! &= \! \int dx \!\int dy \! \left[\frac{\partial }{\partial x} \! \left( \! \ln \frac{p_t(x,y)}{p^X_t (x)p^Y_t (y)}\! \right) \! \right]\! \nu^X_t(x,y) p_t(x, y), \\
    \dot{\mathcal{I}}^Y \! &= \!\int \! dx \! \int \! dy \! \left[\frac{\partial }{\partial y} \! \left( \!  \ln \frac{p_t(x,y)}{p^X_t (x)p^Y_t (y)}\! \right) \! \right]\! \nu^Y_t(x,y) p_t(x, y),
\end{align}
where $\sigma_{{\rm bath}; t}^X$ ($\sigma_{{\rm bath}; t}^Y$) is the entropy change of the system $X$ ($Y$), $\sigma_{{\rm bath}; t}^X$ ($\sigma_{{\rm bath}; t}^Y$) is the entropy change of the heat bath attached to the system $X$ ($Y$), and $ \dot{\mathcal{I}}^X$ ($ \dot{\mathcal{I}}^Y$) is information flow from $X$ to $Y$ ($Y$ to $X$). We remark that this information flow is related to other measure of information flow called the transfer entropy~\cite{ito2013information, ito2016backward}. 

We explain the decomposition of the partial entropy production rates Eqs.~(\ref{decomp}) and (\ref{decomp2}). The entropy changes of the system $X$ and $Y$ are given by the differential entropy change,
\begin{align}
    \sigma_{{\rm sys}; t}^X &=\int dx \frac{\partial  p^X_t (x)}{\partial t}[ -\ln p^X_t (x)]  \notag \\
    &= \frac{d}{dt} S^X_{\rm sys}, \\
    S^X_{\rm sys} &= \int dx [- p^X_t(x) \ln p^X_t (x)], \label{sysen1}\\
    \sigma_{{\rm sys}; t}^Y &= \frac{d}{dt} S^X_{\rm sys},\\
    S^Y_{\rm sys} &= \int dy [- p^Y_t(y) \ln p^Y_t (y)], \label{sysen2}
\end{align}
where we used the partial integral and the normalization of the probability $(d/dt) \int dx p_t^X(x)=0$.
 The sum of the entropy changes of the heat bath gives the total entropy changes of the heat bathes
\begin{align}
    \sigma_{{\rm bath}; t}^X + \sigma_{{\rm bath}; t}^Y &=\frac{1}{ T}\int dx \int dy\frac{\partial  p_t (x,y)}{\partial t} [- V_t(x,y)]  \notag\\
    &=- \frac{1}{T}\frac{dQ}{dt},
\end{align}
where we used the partial integral. The sum of information flows gives the change of the mutual information $I$ between $X$ and $Y$ ,
\begin{align}
    \dot{\mathcal{I}}^X  + \dot{\mathcal{I}}^Y  &=\int dx \int dy \frac{\partial  p_t (x,y)}{\partial t}  \left(\ln \frac{p_t(x,y)}{p^X_t (x)p^Y_t (y)} \right) \notag\\ 
    &=\frac{dI}{dt}, \label{muten}\\
    I &= \int dx \int dy p_t(x,y) \ln \frac{p_t(x,y)}{p^X_t(x)p^Y_t(y)},
\end{align}
where we used the partial integral, the marginalization $\int dy p_t(x,y)=p^X_t(x)$ and $\int dx p_t(x,y)=p^Y_t(y)$,  and the normalization of the probability $(d/dt) \int dx p_t^X(x)=0$, $(d/dt) \int dx p_t^Y(x)=0$, and $(d/dt) \int dx dy p_t(x,y)=0$. Additionally, we obtain 
\begin{align}
    \sigma_{{\rm sys}; t}^X + \sigma_{{\rm sys}; t}^Y -  \dot{\mathcal{I}}^X  -  \dot{\mathcal{I}}^Y  = \frac{dS_{\rm sys}}{dt},
\end{align}
thus the sum of the partial entropy production rates gives the total entropy production rate. 

The non-negativity of the partial entropy production rates gives the second laws of information thermodynamics for the subsystem~\cite{allahverdyan2009thermodynamic,ito2013information,horowitz2014thermodynamics,hartich2014stochastic,horowitz2014secondlaw,ito2015maxwell},
\begin{align}
   \sigma^X_{{\rm bath};t}+\sigma^X_{{\rm sys};t} &\geq \dot{\mathcal{I}}^X, \label{secondinfothermo1}\\
   \sigma^Y_{{\rm bath};t}+\sigma^Y_{{\rm sys};t} &\geq \dot{\mathcal{I}}^Y, \label{secondinfothermo2}
\end{align}
which implies that the entropy changes of the system and heat bath are bounded by information flow in the presence of the subsystem. The sum of two inequalities
\begin{align}
   &\sigma^X_{{\rm bath};t}+\sigma^X_{{\rm sys};t} - \dot{\mathcal{I}}^X + \sigma^Y_{{\rm bath};t}+\sigma^Y_{{\rm sys};t}- \dot{\mathcal{I}}^Y \geq 0,
\end{align}
gives the second law of thermodynamics for the total system
\begin{align}
   \sigma_{t} &\geq 0.
\end{align}
 These inequalities Eq.~(\ref{secondinfothermo1}) and (\ref{secondinfothermo2}) explain a conversion between information and thermodynamic quantities in the context of the Maxwell's demon. In the study of autonomous demon, the system $Y$ can be considered as the Maxwell's demon, and the system $X$ is regraded as the system where the entropy changes $\sigma^X_{{\rm bath};t}+\sigma^X_{{\rm sys};t}$ can be negative. The second laws of information thermodynamics explains the reason why the entropy changes can be negative. Because of the information flow from the demon to the system $\dot{\mathcal{I}}^X$, the entropy changes can be negative.

Based on the results Eqs.~(\ref{appaent}) and (\ref{syuthou2}), we obtain tighter inequalities compared to the second law of information thermodynamics as follows
\begin{align}
\sigma^X_{{\rm bath};t}+\sigma^X_{{\rm sys};t} &\geq \dot{\mathcal{I}}^X + \lim_{\Delta t\to 0} \frac{\mathcal{W} (p^X_{t+\Delta t}, p^X_t)^2}{\mu T \Delta t^2} \geq \dot{\mathcal{I}}^X,\label{ineqinfo1} \\
\sigma^Y_{{\rm bath};t}+\sigma^Y_{{\rm sys};t} &\geq \dot{\mathcal{I}}^Y + \lim_{\Delta t\to 0} \frac{\mathcal{W} (p^Y_{t+\Delta t}, p^Y_t)^2}{\mu T \Delta t^2} \geq \dot{\mathcal{I}}^Y. \label{ineqinfo2}
\end{align}
Thus, the entropy changes of the system and heat bath are tightly bounded by both information flow and the $L^2$-Wasserstein distance. 

We now consider the situation $\sigma^{\rm rot}_t =0$. 
Because the sum of the partial entropy production rates gives the total entropy production rate, the sum of two tighter inequalities gives non-negativity of a measure $I^{\mathcal{W}}$,
\begin{align}
&\lim_{\Delta t\to 0} \frac{I^{\mathcal{W}}}{\Delta t^2} \geq 0, \\
&I^{\mathcal{W}} = \mathcal{W}(p_{t + \Delta t} , p_t)^2 -\mathcal{W}(p^X_{t + \Delta t} , p^X_t)^2 - \mathcal{W}(p^Y_{t + \Delta t} , p^Y_t)^2 .
\end{align}
The equality holds when 
\begin{align}
\nu^X_t(x,y) = \bar{\nu}^X_t(x), \: \: \nu^Y_t(x,y) = \bar{\nu}^Y_t(y),
\label{conditionequality}
\end{align}
because 
\begin{align}
 \sigma^X_t+\sigma^Y_t = \sigma_t = \lim_{\Delta t\to 0} \frac{\mathcal{W}(p_{t + \Delta t}, p_t)^2}{\mu T \Delta t^2},
\end{align}
holds in case of $\sigma^{\rm rot}_t =0$, and we can obtain
\begin{align}
\sigma^X_t = \bar{\sigma}^X_t =\lim_{\Delta t\to 0} \frac{\mathcal{W}(p^X_{t + \Delta t}, p^X_t)^2}{\mu T \Delta t^2}, \\
\sigma^Y_t = \bar{\sigma}^Y_t =\lim_{\Delta t\to 0} \frac{\mathcal{W}(p^Y_{t + \Delta t}, p^Y_t)^2}{\mu T \Delta t^2},
\end{align}
under the condition Eq.~(\ref{conditionequality}). 
The measure $I^{\mathcal{W}}$ quantifies both the statistical independence and the independence of the potential, while the mutual information $I$ only quantifies the statistical independence. Thus, $I^{\mathcal{W}}$ could be an interesting measure of the independence between two systems when stochastic dynamics of two systems are driven by the Fokker-Planck equation. Its non-negativity is decomposed by tighter inequalities of information thermodynamics Eqs.~(\ref{ineqinfo1}) and (\ref{ineqinfo2}).

\section{Example}
\subsection{Stochastic heat engine and geometrical bounds on efficiency}
\begin{figure}
\centering
        \includegraphics[width=\hsize]{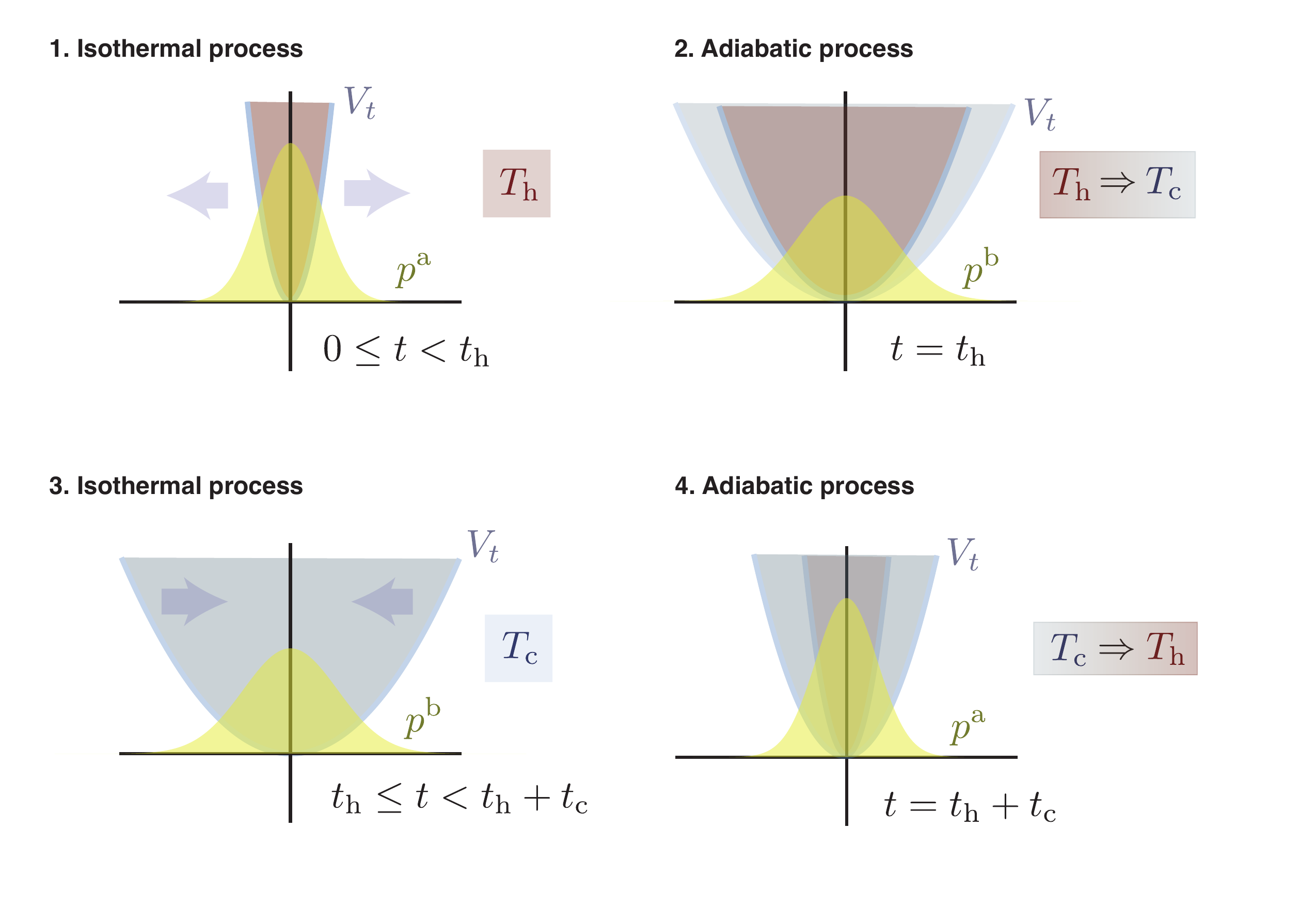}
        \caption{An example of a stochastic heat engine. Because the initial state at time $t=0$ is the same as the final state at time $t=t_{\rm h}+t_{\rm c}$, the four steps gives the cycle of a stochastic heat engine. The work $-W_{\rm h}$ is extracted during time $0 \leq t < t_{\rm h} $, and the work $W_{\rm c}$ is done during time $t_{\rm h} \leq t < t_{\rm h} + t_{\rm c}$. The total amount of the work through one cycle $-W= -W_{\rm h} + W_{\rm c} >0$ is extracted.}
        \label{fig3}
\end{figure}

Let us consider a stochastic heat engine~\cite{Schmiedl2007stochastic} driven by the potential $V_t$. The cycle of a stochastic engine consists of the following four steps (see also Fig.~\ref{fig3}).

\begin{enumerate}
    \item  Let us consider an isothermal process of varying the potential $V_t({\bf x})$ during time $0 \leq t < t_{\rm h} $ at temperature $T_{\rm h}$. During this step, the probability distribution changes from $p^{\rm a}$ to $p^{\rm b}$, and the entropy change of the system is given by $\Delta S := \int d{\bf x} p^{\rm a} ({\bf x}) \ln p^{\rm a} ({\bf x}) - \int d{\bf x} p^{\rm b} ({\bf x})\ln p^{\rm b} ({\bf x})$.
    In this step, the work is extracted $- W_{\rm h} := \int_0^{t_{\rm h}} dt (dW/dt) > 0$ for the external system.

    \item The temperature is changed from $T_{\rm h}$ to $T_{\rm c} (<T_{\rm h})$ instantaneously at time $t=t_{\rm h}$. During this time, the distribution  $p^{\rm b}$ does not change. Therefore, the entropy of the system also did not change, and this step can be interpreted as an adiabatic process.

    \item Let us consider an isothermal process that returns the potential $V_{t_{\rm h}}({\bf x})$ to $V_0({\bf x})=V_{t_{\rm h}+t_{\rm c}}({\bf x})$ during time $t_{\rm h} \leq t < t_{\rm h} + t_{\rm c}$ at temperature $T_{\rm c}$. During this step, the probability distribution changes from $p^{\rm b}$ to $p^{\rm a}$, and the entropy change of the system is $-\Delta S$. 
    In this step, the system is assumed to be given work $W_{\rm c} := \int_{t_{\rm h}}^{t_{\rm h}+t_{\rm c}} dt (dW/dt)  > 0$ by the external system.

    \item The temperature is changed from $T_{\rm c}$ to $T_{\rm h}$ instantaneously at time $t=t_{\rm h}+t_{\rm c}$. During this time, the distribution does not change.
    Therefore, the entropy of the system also did not change, and this step can be interpreted as an adiabatic process.
\end{enumerate}
If we consider the harmonic potential and the initial distribution $p^{\rm a}$ is Gaussian, thermodynamic quantities such as the entropy change and the work are calculated, and we can find an optimal protocol to minimize the entropy production can be obtained analytically~\cite{Schmiedl2007stochastic}. As we shown in the section~V B, $\sigma^{\rm rot}_t=0$ and the entropy production rate is proportional to the action.  

Here we consider a general case that the potential is not necessarily harmonic and the probability distribution at time $t$ is not necessarily Gaussian.
When the time $t_{\rm h}$ and $t_{\rm c}$ are long enough and the potential $V_t({\bf x})$ is a harmonic potential, 
the efficiency of the heat engine becomes the Carnot efficiency asymptotically, and the heat engine can be considered as a stochastic extension of the Carnot cycle.
The extracted work of the heat engine through the one cycle is
\begin{align}
   - W := W_{\rm h} -W_{\rm c} = (T_{\rm h} -T_{\rm c})\Delta S - T_{\rm h} \Sigma_{\rm h} -T_{\rm c} \Sigma _{\rm c}, \label{ineqWork}
\end{align}
where $\Sigma_{\rm h} := \int_0^{t_{\rm h}} dt \sigma_t$  is entropy production in the isothermal step $1$ at temperature $T_{\rm h}$ and $\Sigma_{\rm c}:= \int_{t_{\rm h}}^{t_{\rm h}+t_{\rm c}} dt \sigma_t$ is the entropy production in the isothermal step $3$ at temperature $T_c$. If we assumed that the extracted work is positive $-W>0$, the condition $\Delta S \geq 0$ should be needed because of the second law of thermodynamics $\Sigma_{\rm h}\geq 0$ and $\Sigma_{\rm c} \geq 0$.

By using Eq.~($\ref{oldInequality}$), we can obtain the following inequality for the extracted work $-W$,
\begin{align}
    -W &\leq (T_{\rm h} -T_{\rm c})\Delta S - \frac{\mathcal{W}(p^{\rm a},p^{\rm b})^2}{\mu t_{\rm r}},
    \label{ineqengine}\\
    \frac{1}{t_{\rm r}}&:= \frac{1}{t_{\rm h}}+\frac{1}{t_{\rm c}},
\end{align}
where $t_{\rm r}$ is called the reduced time.  When we impose the positive extracted work in the whole cycle, i.e., $-W>0$, we obtain the following inequality for the reduced time $t_{\rm r}$ from Eq.($\ref{ineqWork}$),
\begin{align}
    \frac{1}{t_{\rm r}} \leq  \mu (T_{\rm h} -T_{\rm c})\frac{\Delta S}{\mathcal{W}(p^{\rm a},p^{\rm b})^2}.
\end{align}
This inequality implies that the reduced time in the engine is generally bounded by the ratio of the square of the $L^2$-Wasserstein distance $\mathcal{W}(p^{\rm a},p^{\rm b})^2$ to the entropy change $\Delta S$, which are given by the initial distribution $p^{\rm a}$ and the final distribution $p_{\rm b}$. 

The efficiency of the heat engine $\eta$ is defined as
\begin{align}
    \eta = \frac{-W}{T_{\rm h} \Delta S -  T_{\rm h} \Sigma_{\rm h}}.
\end{align}
 Because the second law of thermodynamics $\Sigma_{\rm h}+ \Sigma_{\rm c} \geq 0$ holds, we obtain the fact that the efficiency is generally bounded by the Carnot efficiency $\eta_{\rm C}$~\cite{Broeck2013stochastic},
\begin{align}
\eta \leq  \frac{T_{\rm h} -T_{\rm c} }{T_{\rm h}} := \eta_{\rm C}.
\end{align}
When we considered the situation that the entropy production is minimized as follows
\begin{align}
    T_{\rm c} \Sigma_{\rm c}  &= \frac{\mathcal{W}(p^{\rm a},p^{\rm b})^2}{\mu t_{\rm c}},\\
    T_{\rm h} \Sigma_{\rm h} &= \frac{\mathcal{W}(p^{\rm a},p^{\rm b})^2}{\mu t_{\rm h}},
\end{align}
the efficiency $\eta$ is given by 
\begin{align}
    \eta = \frac{T_{\rm h} -T_{\rm c}  - \frac{\mathcal{W}(p^{\rm a},p^{\rm b})^2}{\mu \Delta S t_{\rm r}}}{T_{\rm h}  -  \frac{\mathcal{W}(p^{\rm a},p^{\rm b})^2}{\mu \Delta St_{\rm h}}},
\end{align}
and reaches to the Carnot efficiency $\eta_{\rm C}$ in the limit $t_{\rm h} \to \infty$ and $t_{\rm c} \to \infty$. This fact is also discussed in Ref.~\cite{Fu2021maximal}. In the limit $t_{\rm h} \to \infty$ and $t_{\rm c} \to \infty$. the square of the $L^2$-Wasserstein distance plays the same role as the irreversible ``action" $A_{\rm irr}$ in Ref.~\cite{Schmiedl2007stochastic}.

When $\sigma^{\rm rot}_t =0$, we obtain a geometric interpretation of the efficiency from Eq.~(\ref{syutyou2}),
\begin{align}
    \eta &= \frac{T_{\rm h} -T_{\rm c}  - \frac{1}{\mu \Delta S}\int_0^{t_{\rm h}+t_{\rm c}} dt \left(\frac{d \mathcal{L}_t}{dt} \right)^2}{T_{\rm h}  -  \frac{1}{\mu \Delta S}\int_0^{t_{\rm h}} dt \left(\frac{d \mathcal{L}_t}{dt} \right)^2}.
    \label{geoefficiency}
\end{align}
In this case, we obtain a lower bound on the efficiency
\begin{align}
     \eta_{\rm C}-  \frac{2 \mathcal{C}}{\mu \Delta S T_{\rm h}} \leq \eta \leq \eta_{\rm C},
     \label{actionbound}
\end{align}
where $\mathcal{C}= (1/2)\int_0^{t_{\rm h}+t_{\rm c}} dt (d \mathcal{L}_t/ dt)^2$ is the action measured by the $L^2$-Wasserstein distance. The efficiency $\eta$ can reach to the Carnot efficiency $\eta_{\rm C}$ when the ratio between the action and the Shannon entropy change $\mathcal{C}/\Delta S$ converges to zero. In general $\sigma^{\rm rot}_t \neq 0$ and this lower bound Eq.~(\ref{actionbound}) is generally violated, especially when the non-potential force exists. Thus, the quantity $\sigma^{\rm rot}_t$ might play an important role in a stochastic heat engine with the non-potential force.

\subsection{Analytical calculation of geometric optimal protocol}
We here discuss dynamics of a Brownian particle in a harmonic potential as an example of stochastic thermodynamics based on $L^2$-Wasserstein distance. In this case, we can show $\sigma^{\rm rot}_t=0$, and obtain the protocol of minimizing the entropy production analytically. In terms of the Langevin equation, the time evolution of the position $x(t)$ at time $t$ is given by
\begin{align}
    \frac{dx(t)}{dt} = - \mu \frac{\partial V_t (x) }{\partial x} + \sqrt{2\mu T} \xi(t),
    \label{brownian1}
\end{align}
with the harmonic potential
\begin{align}
    V_t(x) = \frac{1}{2}k_t(x-a_t)^2,
    \label{brownian2}
\end{align}
where $ \xi (t)$ is the Gaussian noise with the mean $\langle \xi (t) \rangle = 0$ and the variance $\langle \xi (t) \xi (t')\rangle = \delta(t-t')$. 
This Langevin equation corresponds to the following Fokker-Planck equation~\cite{Risken1986FokkerPlanck},
\begin{align}
    \frac{\partial p_t(x)}{\partial t} &= -\frac{\partial }{\partial x} (\nu_t (x)p_t (x)), \\
   \nu_t (x) :&= - \mu \frac{\partial }{\partial x}[ V_t (x)  + T \ln p_t(x) ].
   \label{fokker-planck2}
\end{align}
We now assume that the probability distribution at the initial time is Gaussian. For the harmonic potential, the probability distribution at time $t$ is Gaussian if the probability distribution at the initial time is Gaussian. The probability distribution $p_t(x)$ is written as the Gaussian distribution with the mean ${\rm E}[x]_t$ and the variance ${\rm Var}[x]_t$ at time $t$,
\begin{align}
    p_t(x) &= \frac{1}{\sqrt{2\pi{\rm Var}[x]_t}}\exp \left(- \frac{(x-{\rm E}[x]_t)^2}{2{\rm Var}[x]_t} \right),\\
    {\rm E}[x]_t &= \int dx x p_t (x), \\
    {\rm Var}[x]_t &= \int dx x^2 p_t (x) - ({\rm E}[x]_t)^2.
\end{align}
For this Fokker-Planck equation, the time evolution of ${\rm E}[x]_t$ and ${\rm Var}[x]_t$ is given by 
\begin{align}
    \frac{d}{d t}{\rm E}[x]_t &= \mu k_t (a_t - {\rm E}[x]_t),\label{time_derivetivea}\\
    \frac{d}{d t} {\rm Var}[x]_t &= -2 \mu \left(k_t {\rm Var}[x]_t- T \right). \label{time_derivetiveb}
\end{align}
Therefore, the mean local velocity $\nu_t (x)$ is analytically calculated as
\begin{align}
   \nu_t (x) &= -\mu k_t ({\rm E}[x]_t -a_t) +\left( \frac{\mu T}{{\rm Var}[x]_t} -\mu k_t\right) (x-{\rm E}[x]_t),
\end{align}
and the entropy production rate is also calculated as 
\begin{align}
   \sigma_t &= \frac{1}{\mu T}\int dx |\nu_t (x)|^2 p_t(x)\\
  &= \frac{\mu}{T} \left \{ \left( k_t -\frac{T}{{\rm Var}[x]_t} \right)^2 {\rm Var}[x]_t + k_t^2({\rm E}[x]_t-a_t)^2 \right \} .
\end{align}

The Wasserstein distance can be concretely calculated for the Gaussian distribution~\cite{Givens1984Wasserstein,Takatsu2012Wasserstein}. For two probability distributions 
\begin{align}
    p^{\rm a}(x) &= \frac{1}{\sqrt{2\pi {\rm Var}[x]^{\rm a}}} \exp \left( - \frac{(x-{\rm E}[x]^{\rm a})^2}{2{\rm Var}[x]^{\rm a}} \right ),
\end{align}
and
\begin{align}
    p^{\rm b}(x) &= \frac{1}{\sqrt{2\pi {\rm Var}[x]^{\rm b}}} \exp \left( - \frac{(x-{\rm E}[x]^{\rm b})^2}{2{\rm Var}[x]^{\rm b}} \right ) ,
\end{align}
the $L^2$-Wasserstein distance can be written as follows
\begin{align}
    \mathcal{W} (p^{\rm a},p^{\rm b})^2 = ({\rm E}[x]^{\rm a} - {\rm E}[x]^{\rm b} )^2 + \left(\! \sqrt{{\rm Var}[x]^{\rm a}}- \sqrt{{\rm Var}[x]^{\rm b}} \!\right)^2. 
\end{align}
This $L^2$-Wasserstein distance is also known as the Fr\'{e}chet distance~\cite{Dowson1982Dowson}.
Using this analytical expression of the $L^2$-Wasserstein distance for two Gaussian distributions, we can confirm $\sigma^{\rm rot} = 0$ in this case as follows
\begin{align}
    \left(\frac{d\mathcal{L}_t}{dt} \right)^2 &= \lim_{\Delta t \rightarrow + 0} \frac{\mathcal{W} (p_t , p_{t + \Delta t})^2}{\Delta t^2} \notag \\
    &= \left( \frac{d{\rm E}[x]_t}{dt} \right)^2 + \left( \frac{d \sqrt{{\rm Var}[x]_t}}{dt}  \right)^2 \notag \\
    &= \mu^2 \left\{\frac{\left( k_t {\rm Var}[x]_t - T \right)^2}{{\rm Var}[x]_t}  + k_t^2({\rm E}[x]_t - a_t)^2 \right\} \notag \\
    &= \mu T \sigma_t,
\end{align}
where we used Eqs.~(\ref{time_derivetivea}) and (\ref{time_derivetiveb}).

We also can confirm that the entropy production $\Sigma$ is minimized if Eq.~(\ref{minimization}) holds. The minimum value of the entropy production $\Sigma$ for fixed $p_0$ and $p_{\tau}$ is calculated as
\begin{align}
\Sigma &= \frac{\int_0^{\tau} dt \left[ \left( \frac{d{\rm E}[x]_t}{dt} \right)^2 + \left( \frac{d \sqrt{{\rm Var}[x]_t}}{dt}  \right)^2 \right]}{\mu T}  \\
&\geq \frac{\left(\int_{t=0}^{t=\tau} d{\rm E}[x]_t \right)^2 + \left( \int_{t=0}^{t=\tau} d \sqrt{{\rm Var}[x]_t} \right)^2}{\mu T \tau}  \\
&=\frac{\left({\rm E}[x]_{\tau} - {\rm E}[x]_{0}  \right)^2 + \left( \sqrt{{\rm Var}[x]_{\tau}} -\sqrt{{\rm Var}[x]_0} \right)^2 }{\mu T \tau},
\end{align}
where we used the Cauchy-Schwarz inequality $\tau \int_0^\tau dt (d\theta_t/dt)^2 \geq (\int_0^\tau dt (d\theta_t/dt))^2$ with $\theta_t={\rm E}[x]_t$ and $\theta_t=\sqrt{{\rm Var}[x]_t}$. The minimum value can be achieved if $d\theta_t/dt$ is constant. This condition of the minimum value can be rewritten as
\begin{align}
    &{\rm E}[x]_t = \left(1- \frac{t}{\tau} \right) {\rm E}[x]_0 + \frac{t}{\tau} {\rm E}[x]_{\tau}  \\
    &\sqrt{{\rm Var}[x]_t} = \left(1- \frac{t}{\tau} \right) \sqrt{ {\rm Var}[x]_0 } + \frac{t}{\tau}\sqrt{{\rm Var}[x]_{\tau}},  \label{optimal_protocol_of_variance}
\end{align}
or equivalently,
\begin{align}
    & \frac{d{\rm E}[x]_t}{dt} =  \frac{ {\rm E}[x]_{\tau} - {\rm E}[x]_0}{\tau}, \label{optimal_protocola} \\
    & \frac{d\sqrt{{\rm Var}[x]_t}}{dt} = \frac{ \sqrt{ {\rm Var}[x]_\tau } -  \sqrt{ {\rm Var}[x]_0 } }{\tau}. 
    \label{optimal_protocolb}
\end{align}
Under this condition, $\mathcal{W} (p_0, p_{\tau})/\tau$ is calculated as 
\begin{align}
    &\frac{\mathcal{W} (p_0, p_{\tau})}{\tau} \notag \\
    &= \frac{1}{\tau}\sqrt{({\rm E}[x]_{\tau} - {\rm E}[x]_0 )^2 + \left( \sqrt{{\rm Var}[x]_{\tau}}-  \sqrt{{\rm Var}[x]_0} \right)^2} \notag\\
    &=\sqrt{\left( \frac{d{\rm E}[x]_t}{dt} \right)^2 + \left( \frac{d \sqrt{{\rm Var}[x]_t}}{dt}  \right)^2 } \notag \\
    &=  \frac{d \mathcal{L}_t}{dt} ,
\end{align}
which is the condition that the probability distribution changes at a constant rate on a straight line measured by the $L^2$-Wasserstein distance Eq.~(\ref{minimization}). By comparing Eqs.~(\ref{optimal_protocola}) and (\ref{optimal_protocolb}) with Eqs.~(\ref{time_derivetivea}) and (\ref{time_derivetiveb}), the optimal protocol that minimizes the entropy production is given by
\begin{align}
   \mu k_t (a_t - {\rm E}[x]_t) &=\frac{ {\rm E}[x]_{\tau} - {\rm E}[x]_0}{\tau},\\
   - \mu \left(k_t {\rm Var}[x]_t- T \right) &=\sqrt{{\rm Var}[x]_t} \frac{ \sqrt{ {\rm Var}[x]_\tau } -  \sqrt{ {\rm Var}[x]_0 } }{\tau}.
\end{align}
In terms of the parameters of the harmonic potential $V_t(x)$, the optimal protocol that minimizes the entropy production is given by
\begin{align}
   k_t &= \frac{T}{{\rm Var}[x]_t} - \frac{ \sqrt{ {\rm Var}[x]_\tau } -  \sqrt{ {\rm Var}[x]_0 } }{ \mu\tau \sqrt{{\rm Var}[x]_t} }, \label{optharmonica}\\
   a_t &= {\rm E}[x]_{t} + \frac{ {\rm E}[x]_{\tau} - {\rm E}[x]_0}{k_t \mu \tau}.
   \label{optharmonicb}
\end{align}
Thus, we obtain the parameters $k_t$ and $a_t$ which realize such an optimal protocol in practice
\begin{align}
   k_t =& \frac{T \tau^2}{\left(\tau \sqrt{ {\rm Var}[x]_0}  + t (  \sqrt{ {\rm Var}[x]_\tau } -  \sqrt{ {\rm Var}[x]_0 }) \right)^2} \notag\\
   &- \frac{ \sqrt{ {\rm Var}[x]_\tau } -  \sqrt{ {\rm Var}[x]_0 } }{ \mu \left[ \tau \sqrt{ {\rm Var}[x]_0}  + t (  \sqrt{ {\rm Var}[x]_\tau } -  \sqrt{ {\rm Var}[x]_0 } ) \right] }, \label{optimizationk} \\
   a_t =&  \frac{\tau {\rm E}[x]_0  +t ({\rm E}[x]_{\tau} -{\rm E}[x]_0 )}{\tau}+ \frac{ {\rm E}[x]_{\tau} - {\rm E}[x]_0}{k_t \mu \tau}. \label{optimizationa}
\end{align}
If we assume that $k_t$ is always non-negative, the following inequality
\begin{align}
    \tau  & \geq \frac{1- t \mu T}{\mu T}\frac{ \sqrt{ {\rm Var}[x]_\tau } -  \sqrt{ {\rm Var}[x]_0 } }{ \sqrt{ {\rm Var}[x]_0 }  } \\
    & \geq \frac{1}{\mu T}\frac{ \sqrt{ {\rm Var}[x]_\tau } -  \sqrt{ {\rm Var}[x]_0 } }{ \sqrt{ {\rm Var}[x]_0 }  }.
    \label{conditionoptimal}
\end{align}
must hold for this optimal protocol. The results implies that when the variance gets smaller, i.e., ${\rm Var}[x]_\tau < {\rm Var}[x]_0$, we can use this optimal protocol for all $\tau >0$, but when the variance gets larger, i.e., ${\rm Var}[x]_\tau \geq {\rm Var}[x]_0$, there is a limit to the time $\tau$ for the process to achieve this optimal protocol. 

\subsection{Numerical illustration of thermodynamic speed limits}
\begin{figure}
\centering
        \includegraphics[width=\hsize]{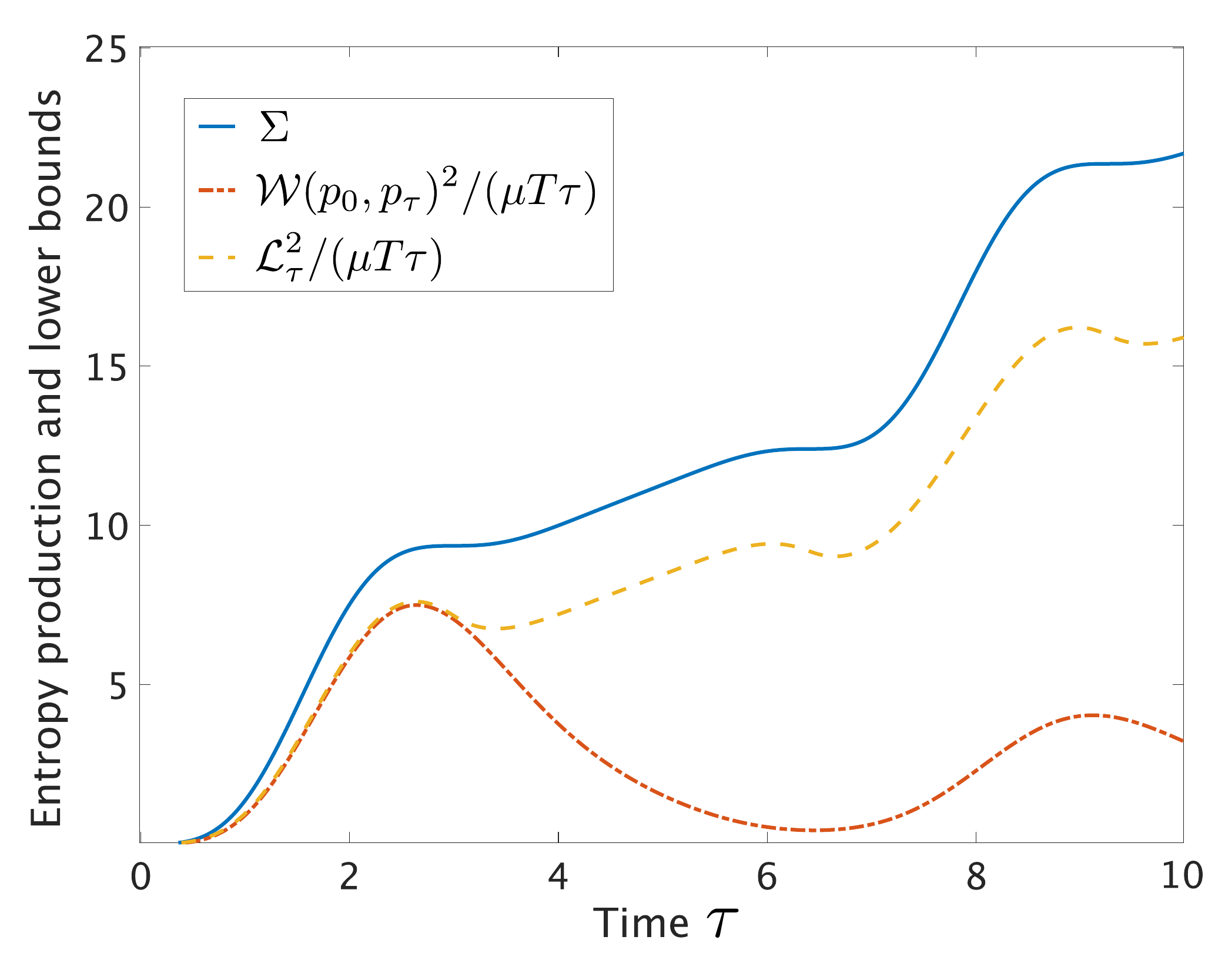}
        \caption{Comparison between two lower bounds on the entropy production $\Sigma$. The lower bound by the Wasserstein path length $\mathcal{L}_{\tau}^2 /(\mu T \tau)$ is much tighter than the lower bound by the Wasserstein distance $\mathcal{W}(p_0, p_{\tau}) ^2 /(\mu T \tau)$.}
        \label{fig4}
\end{figure}
We numerically test lower bounds on the entropy production. We consider the Brownian motion with the harmonic potential Eqs.(\ref{brownian1}) and (\ref{brownian2}). The parameters of the Brownian motion are given by $\mu = 0.01$ and $T = 1$. We consider the case that the parameters of the harmonic potential are periodically changed as follows,
\begin{align}
    k_t &= 2 + \sin(t),\\
    a_t &= 10 \sin(t).
\end{align}
The initial distribution is Gaussian with ${\rm E}[x]_0= {\rm Var}[x]_0=1$, and calculate the time evolution from $\tau =0$ to $\tau =10$. In Fig.~\ref{fig4}, we illustrate the tightness of the lower bound $\mathcal{L}_{\tau}^2 /(\mu T \tau)$ in Eq.~(\ref{newInequality}) compared with the lower bound $\mathcal{W}(p_0, p_{\tau}) ^2 /(\mu T \tau)$ in Eq.~(\ref{oldInequality}). The value of the Wassserstein distance $\mathcal{W}(p_0, p_{\tau})$ oscillates for the periodic change of the potential whereas the value of the Wassserstein path length $\mathcal{L}_{\tau}$ monotonically increases in time. This fact is a reason why our new bound $\mathcal{L}_{\tau}^2 /(\mu T \tau)$ becomes much tighter than the previous bound $\mathcal{W}(p_0, p_{\tau}) ^2 /(\mu T \tau)$ for the periodic change of the potential.

\begin{figure}
\centering
        \includegraphics[width=\hsize]{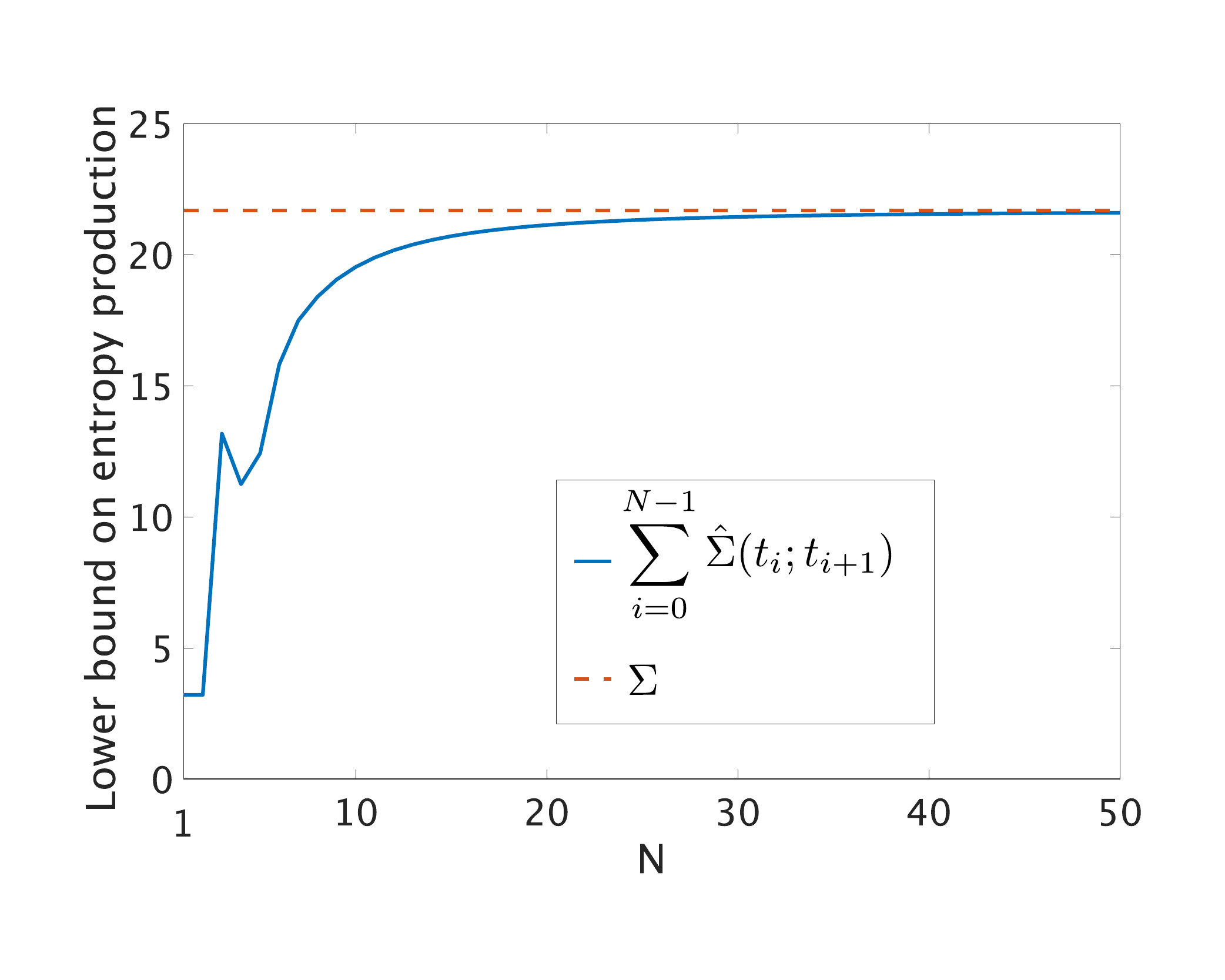}
        \caption{The estimation of the entropy production. The lower bound on the entropy production $\sum_{i=0}^{N-1} \hat{\Sigma}(t_i;t_{i+1})$ converges to the entropy production $\Sigma$. }
        \label{fig5}
\end{figure}

We also numerically check the lower bound in Eq.~(\ref{estimationbound}) and the estimation of the entropy production based on Eq.~(\ref{estimationentropyproduction}). In Fig.~\ref{fig5}, we numerically calculate the lower bound $\sum_{i=0}^{N-1} \hat{\Sigma}(t_i;t_{i+1})$ as a function of the integer $N = \tau (i/t_i)$ at time $\tau =10$. The inequality Eq.~(\ref{estimationbound}) holds for any $N$, and the lower bound converges to the entropy production in the limit $N \to \infty$.  This result implies that $\sigma^{\rm rot} =0$ and the entropy production is numerically estimated from the Wasserstein path length. Our lower bound might be useful to estimate the entropy production from the measurement of the probability distribution at the short time interval. 

\subsection{Numerical calculation of optimal stochastic heat engine in finite time}
We numerically discuss the optimal protocol for stochastic heat engine in finite time. We here consider the Brownian motion with the harmonic potential Eqs.(\ref{brownian1}) and (\ref{brownian2}). We set the parameters as $\mu = 0.01$, $t_{\rm h}=t_{\rm c}=100$, $T_{\rm h} =10$ and $T_{\rm c} =1$. Thus, the Carnot efficiency is given by $\eta_{\rm C}= 0.9$. We assume that the probability distribution is Gaussian with zero mean ${\rm E}[x]_t=0$ and variance ${\rm Var}[x]_t$ satisfying ${\rm Var}[x]_0 = {\rm Var}[x]_{t_{\rm h} + t_{\rm c}} =1$ and $ {\rm Var}[x]_{t_{\rm h}} =4$. We easily check that the condition Eq.~(\ref{conditionoptimal}) holds in this case.
From Eqs. (\ref{optimizationk}) and (\ref{optimizationa}), the optimal parameters of the harmonic potential are given by
\begin{align}
   k_t &= \frac{ 100000}{ (100   + t)^2  } - \frac{ 100 }{ 100   + t  }, \\
   a_t &= 0
\end{align}
for $0 \leq t <t_{\rm h}$ and 
\begin{align}
   k_t &= \frac{ 10000 }{ (300   - t)^2  } + \frac{ 100 }{ 300   - t  }, \\
   a_t &= 0
\end{align}
for $t_{\rm h} \leq t \leq  t_{\rm h}+t_{\rm c}$. By considering a similar discussion to derive Eq.~(\ref{estimationbound}), we can obtain the lower bound on the entropy production as follows,
\begin{align}
   \Sigma &\geq \hat{\Sigma},\\
   \hat{\Sigma} &= \left\{
    \begin{alignedat}{2}
        \frac{\mathcal{W}(p_0, p_{\tau})^2}{ \tau \mu T_{\rm h} } & \: \: \: \:  & (0 \leq \tau < t_{\rm h}) \\
        \frac{\mathcal{W}(p_0, p_{t_{\rm h}})^2}{ t_{\rm h} \mu T_{\rm h} } & +  &\frac{\mathcal{W}( p_{t_{\rm h}}, p_{\tau} )^2}{ (\tau -t_{\rm h})  \mu T_{\rm c} }   \: \: \: \:  & (t_{\rm h} \leq \tau \leq  t_{\rm h}+t_{\rm c}).
    \end{alignedat}
    \right.
    \label{lowerboundefficiency}
\end{align}
In Fig.~6, we show that the lower bound Eq.~(\ref{lowerboundefficiency}) is equal to the entropy production in this optimal protocol. We can check that the time derivative of the Wasserstein path length $d \mathcal{L}_{\tau}/dt$ is constant. 
\begin{figure}
\centering
        \includegraphics[width=\hsize]{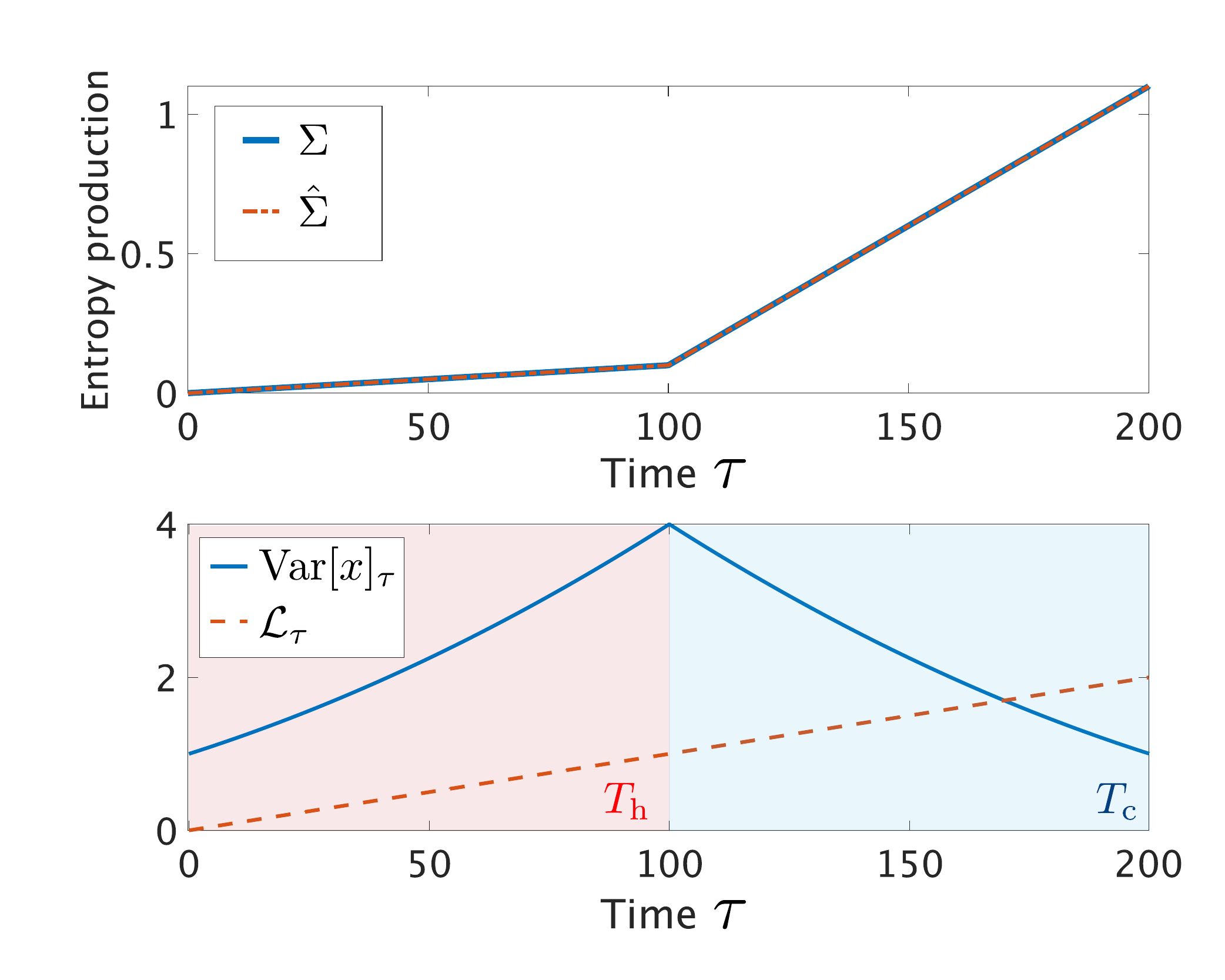}
        \caption{The entropy production $\Sigma$ for the optimal heat engine in finite time. The lower bound on the entropy production $\hat{\Sigma}$ is equal to the entropy production $\Sigma$ in this case.  We also show the time evolution of the variance ${\rm Var}[x]_{\tau}$ and the Wasserstein path length $\mathcal{L}_{\tau}$.
        The red area implies the interval where the temperature of the heat bath is $T_{\rm h}$, and the blue area implies the interval where the temperature of the heat bath is $T_{\rm c}$, respectively.}
        \label{fig6}
\end{figure}

We next discuss the bound on the efficiency in this case. The change of the Shannon entropy is calculated as $\Delta S = \ln 2$ and the Wasserstein distance is calculated as $\mathcal{W}(p^{\rm a}, p^{\rm b}) = 1$. Thus, the efficiency is numerically obtained as $\eta\simeq 0.7145$, which is lower than the Carnot efficiency $\eta_{\rm C}=0.9$. On the other hand, the action is calculated as $\mathcal{C}=0.01$ the lower bound on the efficiency Eq.~(\ref{actionbound}) gives a reasonable value
\begin{align}
\eta_{\rm C} - \frac{2\mathcal{C}}{\mu \Delta S T_{\rm h}} \simeq  0.6115,
\end{align}
which is smaller than the efficiency $\eta \simeq 0.7145$. 

\section{Discussion}
We discuss a geometrical feature of stochastic thermodynamics for the Fokker-Planck equation based on the $L^2$-Wasserstein distance. As shown in this paper, the $L^2$-Wasserstein distance is strongly related to the entropy production for the Fokker-Planck equation. Thus, based on $L^2$-Wasserstein distance, we can introduce a differential geometry of stochastic thermodynamics for the Fokker-Planck equation, closely related to the entropy production.

It might be interesting to consider a relation between the $L^2$-Wasserstein distance and the Fisher information matrix because the Fisher information matrix
gives another metric in information geometry, which is also a possible choice of differential geometry of stochastic thermodynamics. For example, the entropy production is given by the projection in information geometry~\cite{ito2018unified}. Thus, there might be a deep connection between information geometry and optimal transport by the $L^2$-Wasserstein distance. For example, the HWI inequality, the logarithmic Sobolev inequalities, and the Talagrand inequalities are considered as a trade-off relation among the $L^2$-Wasserstein distance, the relative Fisher information, and the Shannon entropy~\cite{Villani2000generalization, Villani2008Optiaml}. As shown in Ref.~\cite{ito2020stochastic}, we have an analogy between the entropy production rate and the Fisher information of time for the Fokker-Planck equation. This analogy is also pointed out in Ref.~\cite{Li2019Wasserstein}. Thus, we might unify two directions of researches of information geometry and the $L^2$-Wasserstein distance for the Fokker-Planck equation based on the entropy production. The unification of information geometry and geometry of the $L^2$-Wasserstein distance has been recently discussed~\cite{amari2018information, amari2019information}, and our results might provide a new direction in this topic.

If we consider thermodynamics based on information geometry, we can consider not only stochastic thermodynamics for the Fokker-Planck equation~\cite{ito2020stochastic} but also stochastic thermodynamics for the Markov jump process~\cite{ito2018stochastic1} and chemical thermodynamics for the rate equation~\cite{yoshimura2020information}. Thus, it might be interesting to seek a correspondence of the $L^2$-Wasserstein distance for the Markov jump process and the rate equation.  Indeed, T. Van Vu and Y. Hasegawa derived a generalization of thermodynamic speed limits for the Markov jump process~\cite{Van2021geometrical}, then a thermodynamic correspondence of the $L^2$-Wasserstein distance for the Markov jump process might be the distance discussed in Ref.~\cite{Van2021geometrical}. Moreover, our result is based on the setting of the overdamped Langevin equation, where the entropy production rate is given by the mean local velocity. Thus, it is interesting to generalize our result for the underdamped Langevin equation or the generalized Langevin equation for a non-Markovian process.

In a nonequilibrium steady state, the quantity $\sigma^{\rm rot}_t$ might play an important role. Under the existence of the non-potential force, the entropy production rate is generally decomposed into two non-negative parts, the Wasserstein part $(d\mathcal{L}_t/dt)^2 /(\mu T)$ and the non-potential part $\sigma^{\rm rot}_t$. This fact is very similar to the case of the steady state thermodynamics~\cite{Hatano2001steady}, where the entropy production is decomposed into the excess entropy production and the housekeeping heat. 

\begin{acknowledgements}
We thank Shun-ichi Amari, Masafumi Oizumi, Kohei Yoshimura, Andreas Dechant, Shin-ichi Sasa, Tan Van Vu and Hideo Higuchi for fruitful discussions.  We also thank Andreas Dechant and Shin-ichi Sasa for critical comments to the previous vesrion of this manuscript.
Sosuke Ito is supported by JSPS KAKENHI Grant No. 19H05796, 21H01560, JST Presto Grant No. JPMJPR18M2 and UTEC-UTokyo FSI Research Grant Program.
\end{acknowledgements}

\appendix
\section{Proof of the formula Eq.~(\ref{theorem})}
To obtain the formula Eq.~(\ref{theorem}), we introduce the map $\mathcal{M}_{t \rightarrow s}$ for the trajectory of the particle according to the Fokker-Planck equation from time $t$ to time $s$. The map $\mathcal{M}_{t\rightarrow t+s}$ is given by the following differential equations for $s \geq 0$,
\begin{align}
    \frac{d}{ds}\mathcal{M}_{t\rightarrow t+s}({\bf x}) &= \bmnu_{t+s} (\mathcal{M}_{t \rightarrow t+s}({\bf x})),
\label{timeevolutionM}
\end{align}
with the initial condition $\mathcal{M}_{t\rightarrow t}({\bf x}) := {\bf x}$. The map $\mathcal{M}_{t\rightarrow t- s}$ for $s \geq 0$ is also given by 
\begin{align}
    \frac{d}{dt}\mathcal{M}_{t\rightarrow t-s}({\bf x}) &= - \bmnu_{t-s} (\mathcal{M}_{t \rightarrow t- s}({\bf x})).
\end{align}
with the initial condition $\mathcal{M}_{t\rightarrow t}({\bf x}) := {\bf x}$. These differential equations correspond to the Lagrangian descriptions of the Fokker-Planck equation as a continuity equation. Because the 
composite map $\mathcal{M}_{t\rightarrow t+s}\circ \mathcal{T}_{t}({\bf x}) =\mathcal{M}_{t\rightarrow t+s} (\mathcal{T}_{t} ({\bf x})) $ is a non-optimal transport plan from $p$ to $p_{t+s}$, we obtain the inequality
\begin{align}
    \mathcal{W}(p,p_{t+s})^2 &= \int d{\bf x} \ \Vert {\bf x}-\mathcal{T}_{t+s}({\bf x}) \Vert^2 p({\bf x}) \nonumber \\
    &\leq \int d{\bf x} \Vert {\bf x}-\mathcal{M}_{t\rightarrow t+s}(\mathcal{T}_{t} ({\bf x})) \Vert^2 p({\bf x}).
    \label{inequalitynonoptimal}
\end{align}
By using Eqs.~(\ref{timeevolutionM}) and (\ref{inequalitynonoptimal}), we obtain
\begin{align}
&\left. \frac{d}{ds}\left(\frac{\mathcal{W}(p, p_{t+s})^2}{2} \right) \right|_{s=0}   \notag \\
    &=\lim_{s\rightarrow +0} \frac{1}{s}\left(\frac{\mathcal{W}(p, p_{t+s})^2}{2} - \frac{\mathcal{W}(p,p_t)^2}{2}\right)  \notag \\
    &\leq  \int d{\bf x} p({\bf x})  \left[ \lim_{s\rightarrow + 0} \frac{\Vert {\bf x} -\mathcal{M}_{t\rightarrow t+s}( \mathcal{T}_t({\bf x}))\Vert^2 - \Vert {\bf x} -\mathcal{T}_t({\bf x})\Vert^2} { 2s } \right]\notag \\
    & = - \int d{\bf x}  ({\bf x}-\mathcal{T}_t({\bf x}) )\cdot \bmnu_t (\mathcal{T}_t({\bf x}))p({\bf x}) .
    \label{ineq1}
\end{align}
Similarly, we obtain 
\begin{align}
&\left. \frac{d}{ds}\left(\frac{\mathcal{W}(p, p_{t+s})^2}{2} \right) \right|_{s=0}   \notag \\
    &= \lim_{s\rightarrow +0} \frac{1}{s}\left(\frac{\mathcal{W}(p, p_{t+s})^2}{2} - \frac{\mathcal{W}(p,p_t)^2}{2}\right)  \notag \\
    &\geq \int d{\bf x}   \! p({\bf x}) \! \left[ \lim_{s\rightarrow + 0} \frac{\Vert {\bf x} \! -  \!\mathcal{T}_{t+s}({\bf x})\Vert^2 \! - \! \Vert {\bf x}  \! -  \!\mathcal{M}_{t+s\rightarrow t}(\mathcal{T}_{t+s}({\bf x}))\Vert^2 }{2s}  \right]\notag \\
    &=-\int d{\bf x} ({\bf x}-\mathcal{T}_t({\bf x})) \cdot \bmnu_t (\mathcal{T}_t({\bf x}))p({\bf x}) ,
     \label{ineq2}
\end{align}
because the
composite map $\mathcal{M}_{t+s \rightarrow t}\circ \mathcal{T}_{t+s}$ is a non-optimal transport plan from $p$ to $p_{t}$.
From Eqs.~(\ref{ineq1}) and (\ref{ineq2}),  we finally obtain the formula Eq.~(\ref{theorem}).

\end{document}